\pdfoutput=1
\documentclass[journal,review,submit,pdftex,oneauthor]{Definitions/mdpi} 

\usepackage{aas_macros}
\usepackage{threeparttable}   
\usepackage{caption}          
\usepackage{rotating} 
\usepackage{booktabs}
\usepackage{tabularx} 
\usepackage{pdflscape}
\usepackage{titlesec}
\usepackage{graphicx} 
\DeclareGraphicsExtensions{.pdf,.png,.jpg}
\newcommand{\chandra}{{\em Chandra}}
\newcommand{\xmm}{{\em XMM-Newton} }
\newcommand{\swift}{{\em Swift-XRT} }
\newcommand{\nustar}{{\em NuSTAR} }
\newcommand{\ep}{{\em Einstein Probe}}
\newcommand{\ergcms}{erg\,cm$^{-2}$\,s$^{-1}$} 
\newcommand{\ergs}{erg\,s$^{-1}$}
\newcommand{\ergshz}{erg\,s$^{-1}$\,Hz$^{-1}$}
\newcommand{\ergscmhz}{erg\,s$^{-1}$\,cm$^{-2}$\,Hz$^{-1}$}

\newcommand{\cm}{cm$^{-2}$ }
\newcommand{\msun}{M\ensuremath{_\odot}}
\newcommand{\ml}{M\ensuremath{_\odot\,\mathrm{yr}^{-1}}}
\newcommand{\kms}{\ensuremath{\mathrm{km\,s}^{-1}}}

\titleformat{\paragraph}
{\normalfont\normalsize\bfseries}{\theparagraph}{1em}{}
\titlespacing*{\paragraph}
{0pt}{3.25ex plus 1ex minus .2ex}{1.5ex plus .2ex}

\firstpage{1} 
\pubvolume{1}
\issuenum{1}
\articlenumber{0}
\pubyear{2025}
\copyrightyear{2025}
\datereceived{ } 
\daterevised{ } 
\dateaccepted{ } 
\datepublished{ } 
\hreflink{https://doi.org/} 

\Title{Multiwavelength view of circumstellar interaction in supernovae}

\TitleCitation{Multiwavelength view of circumstellar interaction in supernovae}

\Author{Poonam Chandra $^{1}$\orcidA{0000-0002-0844-6563}}

\address{%
$^{1}$ \quad  National Radio Astronomy Observatory, 520 Edgemont Rd, Charlottesville VA 22903, USA; pchandra@nrao.edu\\}

\abstract{The interaction of post-explosion supernova ejecta with the surrounding circumstellar medium  creates emission across the electromagnetic spectrum. Since the circumstellar medium is created by the mass lost from the  progenitor star, it carries tell-tale signatures of the progenitor.  Consequently, observations and modeling of radiation produced by the interaction in various types of supernovae have provided valuable insights into their progenitors. 
Detailed studies have shown that the interaction in supernovae begins and sustains over various timescales and lengthscales, with differing mass-loss rates in distinct sub-classes. This reveals  diverse progenitor histories for these stellar explosions. This review paper summarizes various supernova subtypes, linking them to stellar death pathways, and presents an updated supernova classification diagram. We then present a multi-wavelength study of circumstellar interaction  in different supernova classes.  We  also present    unpublished \chandra{} X-ray as well as radio observations of a type IIn supernova, SN 2010jl, which allow us to extend its circumstellar interaction studies to about 7 years post-explosion.  The new data indicates that the  extreme mass-loss rate  ($\sim 0.1$\,\ml{}) in SN 2010jl, reported by Chandra et al. (2015),  commenced within the last 300 years before the explosion.  We  summarize the current status of the field and
 argue that via detailed studies of the circumstellar interaction, a.k.a. “Time Machine” technique, one of
 the big mysteries of stellar evolution, i.e., mapping  supernovae progenitors to
 their explosive outcomes can be solved.}

\keyword{supernovae; core collapse; Ia-CSM, circumstellar interaction, radio, X-ray} 

\begin{document}


\section{Supernovae and their classification}

Supernovae (SNe) are among the most energetic phenomena in the universe, marking the explosive demise  of massive stars \cite{arnett1996supernovae}. Stars with initial masses $\sim3- 8$\,\msun{} end their lives as sufficiently massive ($0.9-1.1$\,\msun{}) carbon–oxygen (C-O) white dwarfs (WDs) \cite{Iben1983,Weidemann2000}, which when in a
close binary system, may accrete mass from a companion star until approaching the Chandrasekhar mass limit ($\sim$1.4\,\msun{}), thus leading to  ignition of runaway carbon fusion in its degenerate core resulting in  thermonuclear SNe  \cite{WhelanIben1973, Nomoto1982, IbenTutukov1984, Webbink1984, nomoto2013thermonuclear}.
These SNe are also termed  as Type Ia SNe (SNe Ia). 
Stars with initial masses $\gtrsim 8$\,\msun{}  continue nuclear 
burning through successive stages beyond helium, ultimately reaching 
silicon burning and forming an iron core \cite{Burbidge1957,Arnett1971, 
Woosley1995}. Since iron-group nuclei cannot release energy through 
fusion, the core becomes gravitationally unstable and undergoes 
catastrophic collapse, resulting in the formation of a neutron star or black hole.  The 
ensuing rebound and neutrino-driven processes launch a core-collapse 
supernova (CCSN) explosion 
\cite{Colgate1966,Bethe1985,woosley2002evolution}.

\begin{figure}
\begin{center}
 \includegraphics[width=0.995\textwidth]{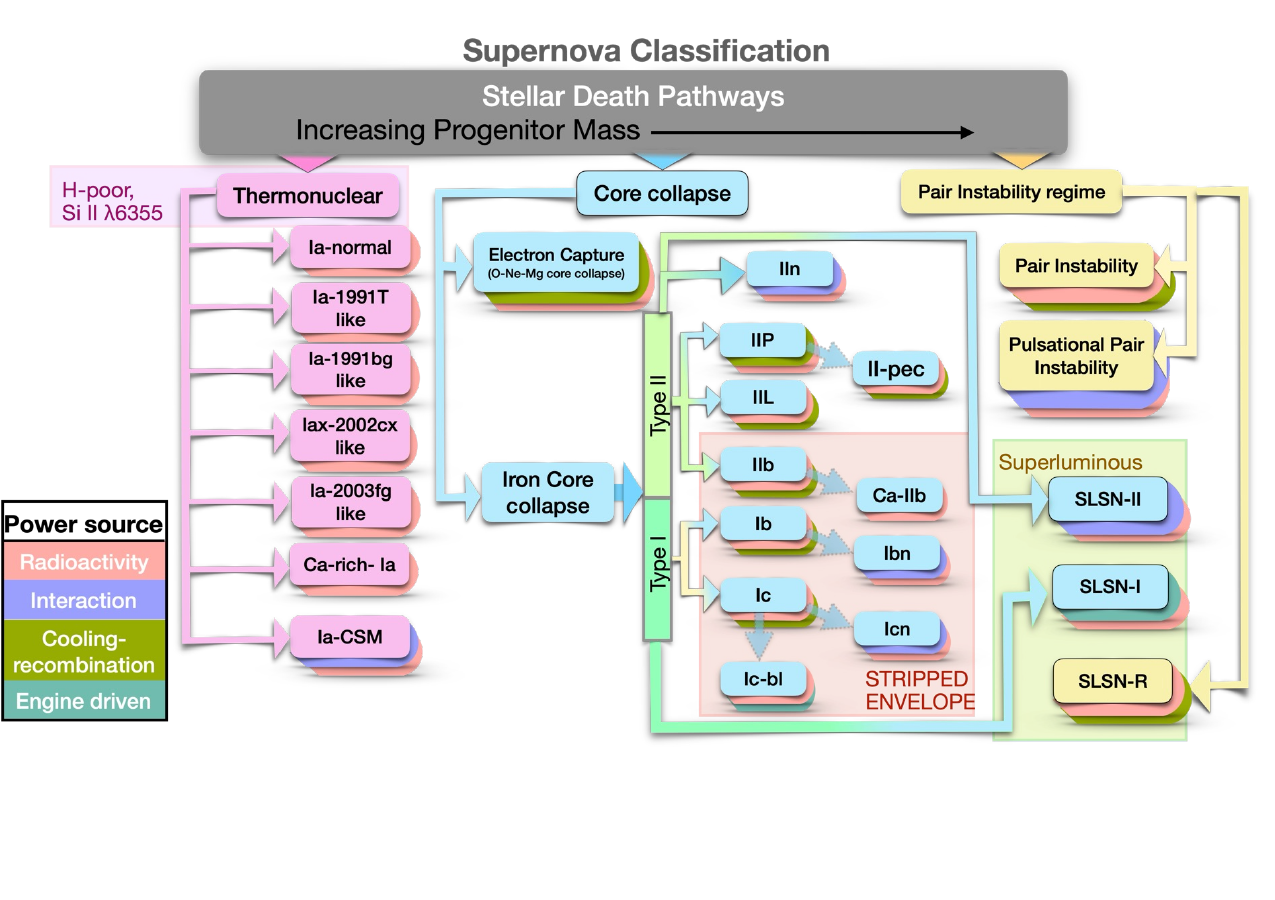} 
 \caption{Supernova classification scheme that combined traditional classification  with stellar death pathways. The two most significant luminosity power sources are mentioned for each subtype.  The color scheme representing the power sources is explained in the  box on the left. }
   \label{fig:SNclassification}
\end{center}
\end{figure} 

SN classification is purely observational, primarily based on optical spectra and light curves. SNe Ia  spectra  are devoid of hydrogen and show the presence of  strong Si\,II\,$\lambda 6355$\,\AA\, lines, e.g.,  SN 1994D,  SN 2011fe \cite{filippenko1997optical, patat1996sn1994d, nugent2011sn2011fe}.
While SNe Ia are expected to be   standardizable candles, observations have revealed significant diversity based on spectral features, light curves, and luminosity \citep{maoz2014observational,taubenberger2017diversity}, such as subtypes 
overluminous  SN 1991T-like SNe Ia \citep{filippenko1992sn1991t,Phillips2024}, and subluminous SN 1991bg-like SNe Ia \citep{filippenko1992sn1991bg}. 
In addition, a peculiar kind of  SNe Ia,  SNe Iax,  are marked by luminosities fainter by a factor 10--100 than normal SNe Ia, slower ejecta speeds, and weaker Si II lines \citep{Foley2013_IaxReview}.  Examples of this subtype are SN 2002cx \cite{Li2003_SN2002cx} and 2005hk \cite{Chornock2006_SN2005hk}, and they are considered to arise from partial deflagration of a WD \cite{Jordan2012_deflagration, Kromer2013_SN2005hkModel}. 
Another category, usually named as SN 2003fg-like SNe Ia,
are extremely bright with absolute peak magnitude $\le -20$ and reveal broad line profiles in their spectra \citep{howell2006sn2003fg,Taubenberger2011}. 

Though  SNe Ia usually do not show hydrogen, some SNe Ia, such as SN 2002ic,  have revealed narrow H$\alpha$ lines, likely arising from  unshocked dense surrounding circumstellar medium (CSM)  \cite{Hamuy2003_2002ic}. They are 
considered to be interacting with H-rich CSM, and are
termed as SNe Ia-CSM. SNe Ia-CSM provides the strongest support for single-degenerate progenitor models \cite{Dilday2012}.  Though SNe Ia-CSM are considered to be thermonuclear explosions, there have been suggestions that some of them may be core collapse events  \cite{Leloudas2015}. Systematic searches, such as using Palomar Transient Factory (PTF) and Bright
Transient Survey of  Zwicky Transient Facility (ZTF) has revealed a population of more than two dozen known SNe Ia-CSM \cite{silverman+13, Sharma+23}.

CCSNe are generally classified based on the presence of hydrogen (Type II) or the absence of hydrogen (Type I) in their spectra
\cite{filippenko1997optical, modjaz2016observational}.
Amongst  type II SNe (SNe II) class, optical lightcurves further subclassify them into Type IIP (SNe IIP) and IIL (SNe IIL) \cite{filippenko1997optical}. SNe IIP show a plateau in the light curve lasting $\sim 100$ days due to hydrogen recombination, and their progenitors are expected to be red supergiant (RSG) stars with a hydrogen envelope intact, e.g., SN 1999em \cite{leonard2002sn1999em}. 
SNe IIL have linearly declining light curves and are expected to come from RSGs with comparatively thinner hydrogen envelopes. A prototypical example of this class is SN 1979C \cite{branch1981sn1979c}.  Modern untargeted surveys reveal a continuum of decline rates between the two traditional subclasses, suggesting the IIP/IIL division is phenomenological rather than being physically discrete  \cite{Anderson2014, Sanders2015}. 
Type I subclass of CCSNe (SNe I) are further divided into Types Ib and Ic (SNe Ib and SNe Ic, respectively).
SNe  Ib show strong He-I lines and are expected to come from helium stars or  Wolf-Rayet (WR) stars, likely in a binary system \cite{Dessart2011,Dessart2020}. 
An   archetype example of the SNe Ib subclass is SN 2008D \cite{modjaz2009sn2008d}. 
Lack of  hydrogen as well as helium is a trademark of SNe Ic, revealing  that their progenitors are  stripped off of both hydrogen and helium layers at the time of explosion (e.g. SN 1994I, \cite{richmond1996sn1994i}). 
In addition, a transitional category between SNe IIL and SNe Ib  has been defined as Type IIb SNe (SNe IIb), which are thought to come from progenitors with partially stripped hydrogen envelopes \cite{Nomoto1993,filippenko1993sn1993j}. SNe IIb, along with SNe Ib and Ic are collectively called stripped envelope SNe (SESNe).  SNe IIP, IIL, IIb, Ib, and Ic are considered to have  progressively increasing stripping of the progenitor off its layers
\cite{filippenko1997optical, GalYam2017}.

Some CCSNe are surrounded by dense CSM and  show narrow emission lines in the optical spectra,  considered to be arising from the dense unshocked CSM \cite{schlegel1990}. They are marked with a suffix "n". Amongst  SNe II, these are SNe IIn \cite{schlegel1990}, and amongst  H-poor CCSNe, these are SNe  Ibn \cite{Pastorello2015} and SNe Icn  \cite{Galyam2022}. They are collectively called interacting SNe. 
 In SNe Ibn, narrow  He I emission lines ($\le1000$\,km\,s$^{-1}$) are  superimposed on a blue continuum \cite{Matheson2004}. Some examples of this class are the first known SNe Ibn
SN 1999cq  \cite{Matheson2004}, and 
a classic well studied SN 2006jc  \cite{Pastorello2007, Foley2007}. 
The first example of SNe Icn was SN 2019hgp, which showed  narrow C III/IV, O III/IV lines in the spectra and a complete lack of H/He \cite{Galyam2022}.
Since interacting SNe are marked by the dominance of the surrounding CSM, their progenitors remain misleading.

An extreme case of SNe Ic is 
SNe Ic-bl.  They  show  broad absorption lines revealing high ejecta velocities.  While most SNe Ic-bl have high but non-relativistic speeds, some have shown relativistic expansion. Some of these are linked with gamma-ray bursts (GRBs), e.g.,  SN 1998bw associated with GRB 980425 \cite{galama1998sn1998bw}, whereas, some relativistic SNe Ic-bl show no GRB association, e.g., SN 2009bb \cite{Soderberg2010}.
SNe Ic-bl  challenge the standard picture of CCSNe due to their large energy requirements. 
A hypothesized scenario includes additional energy being provided by a central engine, such as a relativistic jet driven from the
core of the progenitor or magnetar \cite{Woosley2003}.
However, it is clear that the majority of SNe Ic-bl do not
have GRB counterparts \cite{Corsi2023}.

Electron-capture supernovae (ECSNe) are a theoretical class of faint SNe that come from the core collapse of intermediate-mass  super asymptotic giant branch (super-AGB or SAGB) stars \cite{Poelarends2008}. These stars, unlike their more massive counterparts, are unable to fuse up to iron in their cores \cite{Nomoto1984, Ritossa1999}. Instead, they finish nuclear fusion with a degenerate O-Ne-Mg core. Electron capture onto the Ne and Mg nuclei removes the degenerate electrons, causing a decrease in degeneracy pressure and resulting in a core collapse. ECSNe are predicted to have lower energies and luminosities than their typical, higher-mass, iron CCSNe counterparts. SN 2018zd is by far the best  candidate for an ECSNe \cite{Hiramatsu2021_SN2018zd}.

As mentioned above, the broader SN classification, SNe Ia and CCSNe, is based on optical spectra.  However, as per stellar death pathways, these two classes mainly arise from the explosive death of a WD in a binary system and the gravitational collapse of a massive star, respectively. Another channel of 
stellar death pathways, for stars in the ZAMS range of $140-260$\,\msun{}, is  the pair-instability mechanism \cite{Barkat1967,Heger2002}. This refers to a physical mechanism where a massive star's instability is caused by  high temperatures and pressures in the star's core, leading to the creation of electron-positron pairs. The creation of these pairs reduces the  internal radiation pressure, triggering a runaway thermonuclear explosion \cite{Heger2002}. The resulting SNe are called pair-instability SNe (PISNe).  Stars in the mass
range  $70-140$\,\msun{} undergo 
pulsational pair-instability,  a mechanism where, instead  of a single catastrophic explosion,  a series of violent pulsations  eject large amounts of the star's outer layers.  These are termed as 
pulsational PISNe (PPISNe) \cite{Woosley2017}. SN 2018ibb is considered to be the best  candidate of PISNe \cite{Schulze2024}, whereas the best candidate for PPISNe class is SN iPTF14hls \cite{Arcavi2017}.

Some rare and peculiar subtypes, such as Ca-rich transients, are typically associated with thermonuclear explosions, such as prototypical SN 2005E, considered to be a helium-shell detonation on a low-mass C–O white dwarf \cite{Perets2010,Waldman2011}. However, in the case of another Ca-rich transient,  SN 2005cz, Kawabata et al.  \cite{Kawabata2010} argued that it originated from the core-collapse of a low-mass star. Kasliwal et al.  \cite{Kasliwal2012} carried out a population study showing that Ca-rich transients are faint, Ca-dominated, occur in galaxy halos, and likely come from thermonuclear WD systems.

 A relatively recent  class of SNe,  superluminous SNe (SLSNe), is purely an observational category of SNe whose peak luminosities are much larger than those of
canonical SNe \cite{GalYam2012, Galyam2019, Moriya2024}. They are sub-categorized into Type I and Type II SLSNe, depending upon the absence or presence of hydrogen, respectively. 
Traditionally, the absolute magnitude cut-off for SLSNe has been $M_V<-21$ mag \cite{GalYam2012}, however, some SLSNe I  could be as faint as 
$M_V<-20$ mag and yet have SLSNe spectroscopic features. On the other hand,  some 
fast blue optical transients (FBOTs) may reach $M_V<-21$ mag but do not match with SLSNe spectroscopically, blurring the magnitude boundary between SNe and SLSNe \cite{Galyam2019,Moriya2024}.
SLSNe can come from any stellar death pathways, but three major power sources of  extreme luminosities are radiation from CS interaction, contribution from a central engine (e.g., magnetar spin-down, or fall-back accretion) or radioactive decay in the pair-instability mechanism \cite{Galyam2019,Moriya2024}. 
 Some SNe IIn have  exceptionally high luminosities due to contributions from their dense CSM. These SLSNe IIn are a subset of both SNe IIn and SLSNe, characterized by their narrow hydrogen emission lines and extremely bright light curves, resulting from the efficient conversion of the supernova's kinetic energy into radiation, e.g., SN 2017hcc, SN 2010jl,  SN2006gy, and SN 2014C \cite{Galyam2019}. We discuss them in the SNe IIn category. Among Ic-bl SNe, SN 2011kl associated with GRB 111209A is the only known example of stellar a explosion that satisfies the superluminous criterion, though with faster evolution than usual SLSNe \cite{Greiner2015}. 

 While SN classification typically does not take into account the stellar death pathways, based on the existing understanding, we provide a detailed SNe classification placing the traditional classification along with  the stellar death pathways in Fig. \ref{fig:SNclassification}.
We emphasize that despite various subclasses of SNe, our understanding of their progenitors is quite limited. 
As discussed in the next section,  the SN ejecta-CSM interaction study can help fill this gap in, at least,  a subset of SNe.

\section{Circumstellar interaction}

 Stars lose mass from their outermost layers into the surrounding medium, blowing stellar  winds and creating a CSM.  Observations have shown that mass-loss rates in massive stars can be several orders of magnitude higher than those in average mass stars like our Sun \cite{Puls2008, Smith2014}. If $v_\infty$ is the terminal velocity for a star at a distance of $r_\infty$ and  $\rho(r_\infty)$ is the average mass density  at a distance $r_\infty$, the mass-loss rate can be expressed as $\dot M=4\pi r_\infty^2 \rho(r_\infty)v_\infty$.  The  wind speeds  vary for different progenitors  and are usually related to their escape velocities  \cite{Cassinelli1979ARA&A, Prinja1990, Kudritzki2000}.  However,  the high
radiation from  SN can   accelerate  stellar winds to much higher speeds \cite{Smith2014}. The red super
giants (RSGs) normally have slower winds 
($\sim 10$\,\kms{}) as compared to luminous blue variable (LBV) and blue supergiants (BSGs)  with faster winds ($\sim 100-1000$\,\kms{}). 
P-Cygni profiles  are the best diagnostics of wind speeds \cite{Kudritzki2000}.

There can be multiple  channels of mass loss in massive stars. The major mass-loss mechanisms are  radiation-driven mass-loss, binary interaction-driven mass-loss, and wave-driven mass-loss \cite{Puls2008, Smith2014}. The  radiation-driven winds, such as line-driven, 
continuum-driven and dust-driven winds lead to significant mass-loss, especially during   evolved stages of massive stars \cite{Puls2008, Crowther2007}. Evolved stars like LBVs  are known to be highly 
unstable and can undergo dramatic, violent outbursts resulting in huge amounts of mass-loss \cite{Smith2014}. 
In the second category, massive stars in  binary systems may interact and drive enhanced mass-loss via  Roche-lobe overflow, wind mass transfer, 
common-envelope ejection, and  mergers, etc. processes \cite{Sana2012}. Wave-driven mass-loss
may be important in advanced nuclear burning stages when wave energy generated by deep-seated nuclear burning shells propagates outward and contributes to further enhancement of ongoing mass-loss  by depositing heat and 
momentum into the outer envelopes \cite{Quataert2012,Wu2022A}.  Flash spectra seen in several SNe point towards the prevalence of wave-driven mass-loss very close to explosion \cite{Quataert2012,Wu2021}. In very massive stars ($70 <M<140$\,\msun{}), pulsational pair instability driven mass-loss can be enormous via multiple ejections of  massive shells before the SN collapse \cite{Woosley2017}. As mass-loss from stars forms the CSM, various progenitors can have very different CSM properties. The study of CSM at various time scales carries tell-tale signatures of the SN progenitor stars and their evolution.

\subsection{CSM in massive stars in Milky way}

The best observational evidence of CSM and its complexity is revealed through observations of CSM in massive stars and their evolved counterparts in our Milky Way. Observations of O, early B, and WR stars show CSM created by powerful radiatively driven winds, whereas LBVs, as well as some RSGs, show episodic mass ejections \cite{SMith2011}. 
Massive stars residing in binary systems usually lead to non-spherical CSM \cite{SMith2011}. Several evolved stars in the Milky Way also show CSM ejected via violent outbursts \cite{Meynet2011}.

In Galactic OB stars,  the mass-loss rates are found to be around $10^{-7}-10^{-6}$\,\ml{} along with wind speeds of $1000-3000$\,\kms{} \cite{Puls2008, Smith2014}. Puls \cite{Puls2008} work has shown the formation of infrared (IR) bow shocks at the junction of CSM and interstellar medium (ISM) in some of these stars.  These bow shocks are created as the star is ejected from its birth cluster at high speeds and plows through the ISM, such as $\zeta$ Ophiuchi \cite{Gull1979}. 
Thermal radio emission has been observed from OB stars due to the presence of dense CSM \cite{Wright197,Kurapati2017, Yiu2024}, though a binary companion can lead to non-thermal radio emission via wind collisions \cite{DeBecker2007}. In a subset of massive OB stars hosting kG dipolar magnetic fields \cite{Wade2016}, evidence of gyrosynchrotron radio emission is also observed, showing variability of radio emission w.r.t. rotational phase due to misalignment of magnetic and rotational axes \cite{Linsky1992ApJ}. A small fraction of these magnetic OB stars have also revealed exotic and rare electron cyclotron maser emission,  and  are named as main-sequence radio-pulse emitters (MRPs) \cite{Das2022}. 

WR stars in Milky Way show very dense, optically thick radiatively driven winds with $\dot M\approx 10^{-5}-10^{-4}$\,\ml{} along with wind speeds of $1000-3000$\,\kms{} \cite{Crowther2007, Nugis2000}. 
In some cases, bubble nebulae are seen at a few pcs,  indicators of WR winds sweeping the prior RSG winds. A classic example is  WR 136, along with its complex nebula, NGC 6888
which was created as the star's faster  wind slammed into material ejected during an earlier RSG  phase \cite{Marston1995}.

 Milky Way RSGs  show slow ($\sim10$\,\kms{}) dusty winds with arcs and bows \cite{Georgy2013}. Famous examples of Mily Way RSGs are Betelgeuse and VY CMa.   Betelgeuse is a prime candidate for a CCSN and shows multiple arcs and a bow shock at $\sim6'$ in the IR bands, revealing a complex system of ISM gas and dust shaped by the star's powerful stellar wind colliding with the surrounding medium \cite{Mohamed2012}. 
Betelgeuse experienced a "Great Dimming" between 2019 and 2020, likely caused by a massive ejection of gas that condensed into a temporary dust cloud, further obscuring the star \cite{Jadlovsky2024}.
The RSG VY CMa
shows clumpy, dusty ejecta and asymmetric outflows 
revealing a complex CS environment and pointing towards  erratic mass-loss history as revealed via Atacama Large Millimeter Array (ALMA) mm studies 
\cite{Kaminski2019,Singh2023}.

LBVs are extremely rare, with only a few dozen
known examples in the Milky Way \cite{Mehner2024}. They  show shells, eruptions, and bipolar outflows. $\eta$-Carinae  is the most well-known example in the Milky Way LBVs.
During its "Great Eruption" in the 1840s, $\eta$-Carinae expelled at least 10\,\msun{} of gas and dust, which formed a distinctive, heavily obscured, bipolar Homunculus Nebula \cite{Smith2003}. Other well-known LBVs are P Cygni, AG Carinae, Pistol star, etc. \cite{Groh2006,Figer1998,Gootkin2020}. Most of the LBVs in the Milky Way seem to reside in multiplicity \cite{Mahy2022}.

BSGs are the evolved phase of massive stars, often after the RSG phase \cite{Saio2013}. Arguably the most famous SN of the current time, SN 1987A, had a BSG progenitor \cite{Podsiadlowski2017}. In the Milky Way, 
Sher 25, with its equatorial ring and bipolar outflows, is a well-known example of a BSG, often cited as a SN 1987A progenitor analog \cite{Smartt2002}. A significant number of BSGs are argued to have formed via stellar merger \cite{Menon2024}. 

While the YSG phase is typically before the RSG phase and single stars are not expected to explode in this phase, some SNe, such as SN 2011dh, have been identified to have a YSG progenitor \cite{Maund2011}. However, it has been argued that enhanced mass-loss rate, possibly via binary interaction, can result in explosions \cite{Georgy2012}.
The CSM of YSGs  contains significant amounts of dust and often exhibits episodic or variable mass-loss history, leading to the formation of clumpy, structured CSM \cite{SMith2011}. A famous example of this class is IRC+10420, which shows helium and nitrogen-enriched CSM along with massive dust shells \cite{Humphreys2002}. 

In Table \ref{tab:Milky_Way}, we list some well-known massive and evolved stars in the Milky Way and their CSM properties. As is evident, the mass-loss histories of Galactic massive stars are quite complex. The  resulting CSM has a non-reversible, significant impact on their post-death SN outcome \cite{Chiosi1986}.

\begin{table}
\centering
\scriptsize
\caption{CSM properties of Milky Way Massive stars}
\label{tab:Milky_Way}
\begin{tabular}{lp{2.5cm}llp{2.8cm}p{2cm}l}
\hline
\textbf{Progenitor} & \textbf{mass-loss} & \textbf{Typical $\dot M$} & \textbf{Wind} & \textbf{CSM morphology} & \textbf{Milky Way} & \textbf{Ref.} \\
\textbf{class} & \textbf{mechanism} & \ml{} & \kms{}  &  & \textbf{}  &  \\
\hline
O \& B & Radiation, line-driven & $10^{-8}$-$10^{-6}$  & 1000–3000 & Low-density stellar wind, IR bow shocks if runaway & $\zeta$ Oph; E-BOSS & \cite{Puls2008}\\
WR & Radiatively driven, optically thick & $10^{-5}$-$10^{-4}$ & 1000–3000 & WR bubbles/rings on pc scales & WR 136, WR 124, NGC 2359  & \cite{Crowther2007, Nugis2000}\\
RSGs & Radiation pressure, pulsation/convection & $10^{-6}$-$10^{-4}$& 10–30 & Dusty arcs/bow shocks, clumpy, asymmetric outflows & Betelgeuse, VY CMa  & \cite{Decin2012,Sanders2015}\\
LBVs & Quiescent winds, eruptions, outbursts & $10^{-5}$-$10^{-1}$ & 50–200 & Massive dusty shells, bipolar lobes & $\eta$ Car, AG Car; Pistal star  & \cite{Smith2003, SMith2011}\\
BSGs & RSG winds,  binary driven & $10^{-7}$-$10^{-4}$  &100–2000 & Equatorial ring, bipolar lobes & SBW1, Sher 25 & \cite{Smartt2002, SMith2013MNRAS}\\
YSGs & Pulsation, outbursts & $10^{-5}$-$10^{-3}$ & 35–100 & Multiple shells/arcs; cool dusty CSM & $\rho$ Cas; IRC+10420 & \cite{Humphreys2002,Lobel2003IAUS..210P.F10L}\\
\hline
\end{tabular}
\end{table}

 \subsection{Self-similar solutions for CS interaction}
 
\begin{figure}
\begin{center}
 \includegraphics[width=0.98\textwidth]{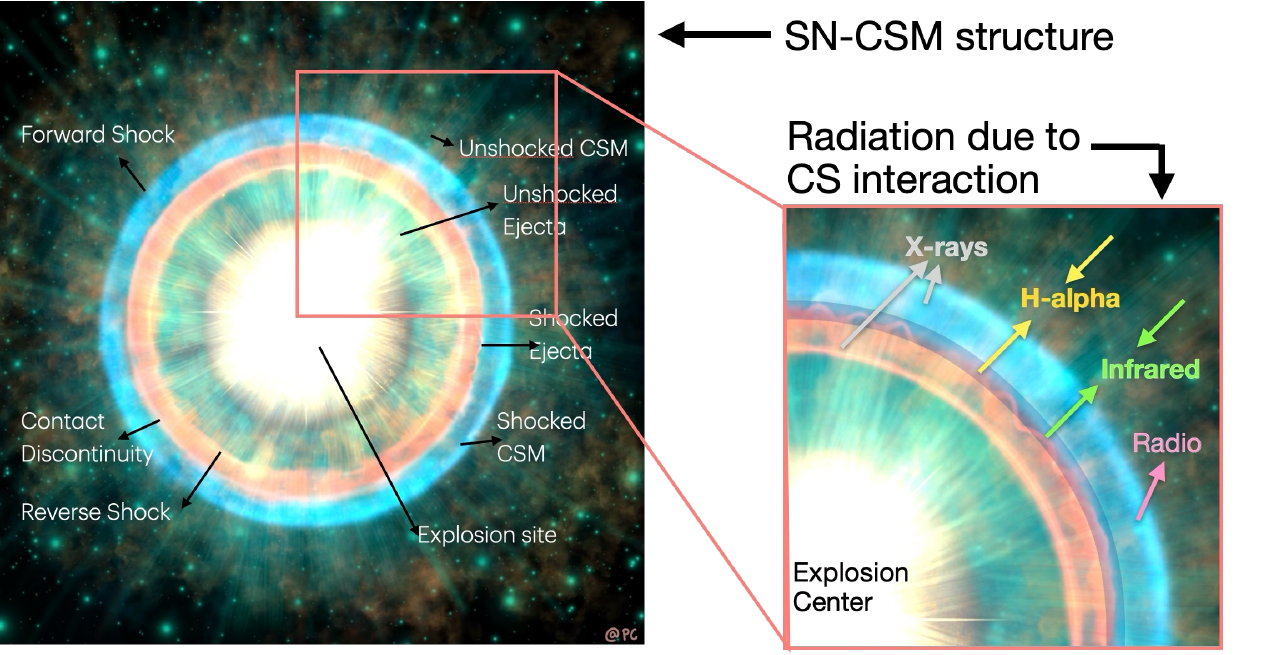} 
 \caption{CS interaction picture in a typical SN. The left image shows ejecta-CSM interaction and the formation of various  shocked and unshocked regions. The right side image shows the radiation produced as a result of ejecta-CS interaction.}
   \label{fig:CSM2}
\end{center}
\end{figure}

In the CS  interaction picture, the rapidly expanding SN ejecta with velocities $\sim 10,000-50,000$\,\kms{}, collides with the surrounding material a.k.a. comparatively slower moving CSM  ($\sim 10-1000$\,\kms{}). This  generates a forward shock (FS) moving into the CSM, a reverse shock  (RS) moving back into the ejecta, separated by a contact discontinuity (CD) where Rayleigh-Taylor instabilities may dominate \cite{Chevalier1982a} (see Fig. \ref{fig:CSM2}). The shocks convert the kinetic energy of the ejecta into thermal and non-thermal energy, producing radiation observable in different wavelengths \cite{Chevalier1982b}.   Hot shocks produce emission in X-ray bands, while electrons accelerated in the shock, in the presence of magnetic fields (enhanced at the CD), produce non-thermal synchrotron radio emission. In optical bands, the CS interaction manifests as narrow lines due to recombination in photoionized unshocked CSM, and as intermediate lines if a cool dense shell (CDS) is formed at the CD between the shocks.
 Infrared (IR) emission may arise as a result of CS Interaction from either the heated dust in the CSM or newly formed dust in the post-shocked regions \cite{Chevalier2003, Chevalier2017} (Fig. \ref{fig:CSM2}).

The fact that the shock velocities are typically 100–1000 times faster than 
CSM winds \cite{Chevalier2003} involve the shock wave sampling the wind lost from the SN progenitor many hundreds to thousands of years ago. It probes the past evolution and mass-loss history of the star during the less advanced nuclear burning stages leading up to the explosion. This property allows CS interaction to be  an important tool to probe the SN progenitor evolution before the explosion, and serves as a  time machine to study various SN progenitors.

The concept of CS interaction was theoretically predicted even before the observational confirmation came \cite{Chevalier1982a}. The CS interaction model  gained acceptance when it successfully explained radio observations in  SN 1979C and SN 1980K  \cite{Weiler1986}. More 
SNe, including SN 1986J and SN 1988Z, provided strong evidence for CS interaction, revealing bright X-rays from hot shocked regions, radio synchrotron emission indicative of electron acceleration, narrow optical lines suggesting ionized CSM, and slowly declining optical light curves that implied an additional energy source beyond the radioactive decay \cite{schlegel1990,weiler1990}. By the 2000s, CS interaction had become the standard framework for explaining radio and X-ray emission from SNe \cite{Smith2007, Chandra2012, Chevalier2017, Chandra2018}.

\begin{figure}
\begin{center}
 \includegraphics[width=0.98\textwidth]{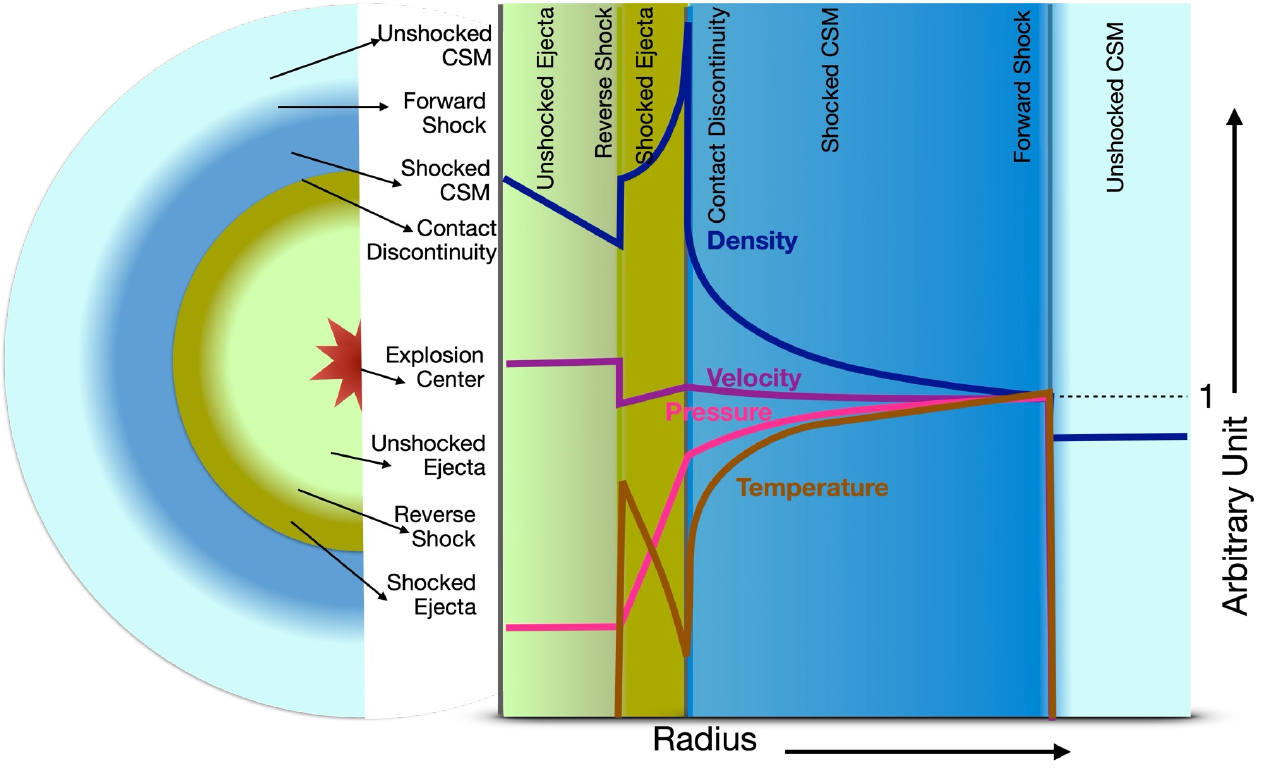} 
 \caption{A cartoon diagram of  variation of density, pressure, and  velocity and temperature in the ejecta-CSM interaction region for $\rho_{\rm ej} \propto r^{-n}$,
 $n\ge5$ and $\rho_{\rm CSM} \propto r^{-2}$. The figure is inspired by the Chevalier model \cite{Chevalier1982a}. The  variables have been normalized to their values at the forward shock front. The quantities are not to be scaled.}
   \label{fig:self-similar}
\end{center}
\end{figure}

Chevalier \cite{Chevalier1982a} first proposed self-similar solutions to model the CS interaction, and it is considered the cornerstone work in the field. In this work, the ejecta-CSM interaction producing  FS and   RS shocks  is governed by the hydrodynamic equations of mass, momentum, and energy conservation (Fig. \ref{fig:self-similar}).  Both ejecta and CSM are assumed to follow power-law density profiles; SN ejecta are defined by $\rho_{\rm ej} \propto r^{-n}$,  with a sharp cutoff at an inner core, interacting with wind like CSM $\rho_{\rm CSM} \propto r^{-2}$. The ratio of RS and radii and the pressure profiles remain fixed in this model. The self-similar solutions provide scaling relations for shock radius, $R_{\rm s}$, evolution with time, $t$, such that $R_{\rm s}\propto t^{\frac{n-3}{n-2}}$. This analytic model provides  a framework to calculate shock velocities, temperatures, and densities analytically, rather than numerically (Fig. \ref{fig:self-similar}).
In  \cite{Chevalier1982b},  Chevalier applied the self-similar hydrodynamical models to the observed radio and X-ray emission seen in some CCSNe, and found the models to be largely successful.
Soon after Chevalier's \cite{Chevalier1982a} analytical self-similar solutions,  Nadyozhin \cite{nadyozhin1985} independently provided numerical evaluation of self-similar solutions. In his models, the  self-similarity  ceased to exist at the CD. This work explored a broad range of density profiles, allowing for the applicability of the model in a wide variety of astrophysical scenarios. 

Hamilton and Sarazin  \cite{hamiltonsarazin1984} also proposed a self-similarity solution, focusing on supernova remnants (SNRs), unlike Chevalier's \cite{Chevalier1982a} model, which was applicable to CS interaction at early times in young SNe. 
Their model, which treated the RS as an infinitely thin and  freely expanding shell, was tested with a uniform ISM and successfully explained X-ray emission from SNRs. It is important to note that while both Chevalier and Nadyozhin's solutions were valid for $n>5$, Hamilton and Sarazin \cite{hamiltonsarazin1984}  derived solutions for flat ejecta $n=0$.

Truelove and McKee  \cite{truelove1999} also developed analytical and numerical solutions for CS interaction, detailing the complete evolutionary sequence of non-radiative SNRs from the ejecta-dominated (free-expansion) to the late time Sedov-Taylor phase (accumulation of significant CSM). Their model offered diagnostic tools for determining explosion energy, ejecta mass, and age, but did not account for Rayleigh-Taylor instabilities at the CD, thermal conduction, or convection.

In 1994, Chevalier and Fransson \cite{Chevalier1994} 
 built upon Chevalier \cite{Chevalier1982a} framework, incorporating radiative cooling in the RS during early phases. They also introduced the concept of CDS, formed at the CD in case of radiative RS, which could lead to mixing of ejecta and the CSM.
While Chevalier's earlier work briefly touched upon radio absorption, a more systematic treatment for absorption of radio emission was presented in his 1998 work
\cite{Chevalier1998}. This laid the groundwork for understanding the synchrotron self-absorption (SSA) caused by synchrotron radiation-emitting electrons and the free-free absorption (FFA) by the external medium. They further proposed that SSA would be dominant in SNe Ib/c with high ejecta velocities \cite{Chevalier2006_Ib}, while FFA would be more significant in SNe II \cite{Chevalier2006_IIP}.

Recently, in 2024, Coughlin  \cite{coughlin2024} expanded upon previous self-similar models, offering solutions that differ from Chevalier's model by allowing the FS, RS, and CD to expand at varying rates, rather than having the same time evolution as in previous models.
Coughlin's models are applicable to an earlier phase than Chevalier's energy-conserving phase, specifically during the initial coasting phase where the ejecta density is considerably higher than the ambient CSM density. These models can also be extended to include relativistic speeds with arbitrary Lorentz factors, making them suitable for phenomena such as GRBs and relativistic SNe.

\subsection{CS interaction formulation}

Chevalier's model  is widely favored for its straightforward application to observations (Fig. \ref{fig:self-similar}). Their work, detailed in Chevalier \& Fransson \cite{Chevalier2003,Chevalier2017}, offers a comprehensive synthesis of over thirty years of research in this area.
Here we reproduce relevant equations, while encouraging the reader to consult the original references for a complete formulation. In the following, 'CS' denotes shocked CSM and 'rev' refers to shocked ejecta.

The ratio of swept up  masses behind the FS ($M_{\rm CS}$) and the RS 
($M_{\rm rev}$), as well as  the ratios of their respective densities can be written as

\begin{eqnarray}
\frac{M_{\rm rev}}{M_{\rm CS}}=\frac{(n-4)}{2} \,\, {\rm and}\,\,
\frac{\rho_{\rm rev}}{\rho_{\rm CS}}=\frac{(n-4)(n-3)}{2} 
\label{density}
\end{eqnarray}

The maximum ejecta velocity  $V_{\rm ej}$,
FS velocity $V_{\rm CS}$, and RS velocity $V_{\rm rev}$
 can be written in terms of shock radius $R_{\rm s}$,

\begin{eqnarray}
V_{\rm ej}=R_{\rm s}/t \propto t^{-1/(n-2)}\\ \nonumber
V_{\rm CS} =dR_{\rm s}/dt=(n-3)/(n-2)V_{\rm ej}\\ \nonumber
V_{\rm rev}=V_{\rm ej}-V_{\rm CS}=V_{\rm ej}/(n-2)
\label{eq:V}
\end{eqnarray}

 Temperatures of the shocks  $T_{\rm CS}$,  $T_{\rm rev}$  (assuming 
cosmic abundances and electron-ion equipartition (neglecting radiative cooling), respectively,   are

\begin{eqnarray}
kT_{\rm CS}=117 \left(\frac{n-3}{n-2}\right)^2 \left(\frac{V_{\rm ej}}{10^4 \,\rm km \, s^{-1}}\right)^2 \rm keV\\ \nonumber
 kT_{\rm rev}=  \frac{T_{\rm CS}}{(n-3)^2}=1.2 \left(\frac{10}{n-2}\right)^2 \left(\frac{V_{\rm ej}}{10^4 \,\rm km \, s^{-1}}\right)^2 \rm keV
 \label{eq:T}
\end{eqnarray}

The hot RS and FS shocks may produce soft and hard X-rays by free-free  bremsstrahlung radiation. 
The total X-ray luminosity is
\begin{equation}
L_{\rm i} \approx 3.0 \times 10^{39} \overline g_{\rm ff} C_{\rm n} \left( \frac{\dot M_{-5}}{v_{\rm wind,1}}\right)^2  \left( \frac{t}{10}\right)^{-1}
 \, \rm erg \, s^{-1}
 \label{eq:xray}
 \end{equation} 
Here $C_n$ is 1 for the FS  and is $(n - 3)(n -  4)^2/4(n - 2)$ for 
the RS. Here   $\dot M_{-5}$ is the mass-loss rate in units $10^{-5}$\,\ml{},
$v_{\rm wind,1}$ is wind speed in 10\,\kms{} and $\overline g_{\rm ff}$ is the free-free Gaunt factor.
This formulation is valid for adiabatic shocks. In the case of radiative RS, a CDS will form at the CD and will absorb  soft X-rays coming from the RS and re-radiate   into UV and optical emission.

 SNe usually remain bright in radio bands for a long time, due to their low energy. Chevalier \cite{Chevalier1998} derived detailed models of synchrotron radio emission incorporating SSA and FFA.
In their model,  SSA and FFA flux densities are, respectively:
\begin{align}
F_{\nu}(t) &= K_{1}\,\nu^{5/2}\,t^{a}\left[1-\exp\!\bigl(-\tau_{\nu}^{\mathrm{SSA}}\bigr)\right],\\ \nonumber
\tau_{\nu}^{\mathrm{SSA}} &= K_{2}\,\nu^{-(p+4)/2}\,t^{-(a+b)},
\end{align}
and 
\begin{align}
F_{\nu}(t) &= K_{1}\,\nu^{-\alpha}\,t^{-\beta}\,
             \exp\!\bigl[-\tau_{\nu}^{\mathrm{FFA}}(t)\bigr],\\ \nonumber
\tau_{\nu}^{\mathrm{FFA}}(t) &= K_{2}\,\nu^{-2.1}\,t^{-\delta},
\end{align}

Here $K_1$ and $K_2$ are the flux density and absorption normalization parameters, respectively; and $\tau_{\nu}^{\mathrm{SSA}}$ and $\tau_{\nu}^{\mathrm{SSA}}$
are SSA and FFA optical depths, respectively. Parameters  $a$, $b$,
$\alpha$, $\beta$, $\delta$ are power law indices. Here $p$
is the electron energy index, which is related to $\alpha$ as $\alpha=(p-1)/2$ and $\delta$ depends upon ejecta density structure as $\delta=(n-3)/(n-2)$.

While FFA provides direct mass-loss rate estimation (for fully ionized CSM with solar abundance), which is 
\begin{equation}
\left( \frac{\dot M_{-5}}{v_{\rm wind,1}}\right)
   = 4.76\,
     \bigl(\tau_{\nu}^{\mathrm{FFA}}\bigr)^{0.5}
     \left(\frac{V_{\mathrm{ej}}}{10^{4}\;\mathrm{km\,s^{-1}}}\right)^{1.5}
     \left(\frac{t}{100\;\mathrm{d}}\right)^{1.5}
     \left(\frac{T_{e}}{10^{5}\;\mathrm{K}}\right)^{0.675}
     \;M_{\odot}\,\mathrm{yr^{-1}}/10\,\rm km\,s^{-1} .
\end{equation}

In the case of SSA, one can obtain the mass-loss rate by relating the post-shock magnetic energy density to the shock ram
pressure, which is
\begin{equation}
\left( \frac{\dot M_{-5}}{v_{\rm wind,1}}\right)
   = \frac{0.06}{m_{H}^{2}}
     \left(\frac{B}{1\;\mathrm{G}}\right)^{2}
     \left(\frac{t}{100\;\mathrm{d}}\right)^{2}
     \;M_{\odot}\,\mathrm{yr^{-1}}/10\,\rm km\,s^{-1},
\end{equation}

Observationally, X-ray telescopes usually provide flux within a certain energy range, whereas radio telescopes typically measure monochromatic radio flux, a.k.a. flux density $F_\nu$. The observed flux density is  expressed in units of \ergscmhz.
 However, due to their often low values, radio flux densities are commonly measured in Jansky (Jy), where 1 Jy is equivalent to 
10$^{-23}$\,\ergscmhz.

\section{Circumstellar interaction in various supernovae types}

In this section, we summarize  observational properties of CS interaction and their interpretation in various SN types.

\subsection{Type Ia supernovae}
SNe Ia typically lack evidence of  CS interaction; however, there have been cases indicating indirect presence of CSM. For instance, SN 2006X  showed variable Na ID absorption lines, indicative of time-variable ionization of CSM gas close to the progenitor \cite{Patat2007_CSM}. A high-resolution survey of about 3 dozen SNe Ia by Sternberg et al \cite{Sternberg2011_CSM} revealed that 20-25\% of SNe Ia in spiral galaxies exhibit time-variable or blueshifted Na ID lines. This indicates the presence of CSM, as absorption due to progenitor winds is expected to be blueshifted relative to the SN.

Mo et al. \cite{Mo+2025} conducted a comprehensive study of  SNe Ia using mid-IR  data from the NEOWISE space telescope, to search for late time mid-IR 
rebrightening and  found 5 out of 8500 SNe Ia showed this signature suggesting the presence of multiple or extended detached CSM shells located at $10^{16}-10^{17}$ cm from the progenitor. These shells contained dust masses $10^{-6}-10^{-4}$\,\msun{}. Their findings indicated that at least 0.05\% SNe Ia exhibit a mid-IR signature of delayed CS interaction, making them more common than previously thought. 
No radio emission was seen in any SN in their sample in the follow-up observations. However, this study underscores the importance of mid-IR observations in unraveling  complex SNe Ia environments.  Terwel et al. \cite{terwel+25} also used $\approx 3600$ ZTF SNe Ia to search for late-time CSM interaction signatures and identified three such cases, which were SNe 2018grt, 2019dlf, and 2020tfc.

\begin{table*}
\centering
\caption{Radio and/or X–ray detections of SNe Ia/Ia-CSM}
\label{tab:radio_xray_IaCSM}
\renewcommand{\arraystretch}{1.15}
\begin{tabular}{lp{3.5cm}cccl}
\hline
SN &  Detection status & Epoch & Luminosity
& $\dot M$ & Ref.\\
& & days & cgs units &  $M_\odot\,\rm yr^{-1}$   & \\
\hline
2005ke & X-rays  with {\it Swift-XRT} (tentative) & 
8--120 &
$L_{0.3-2\,\rm keV}\approx 2\times 10^{38}$  & 
$3\times10^{-6}$ &  \cite{Immler+2006}\\
2012ca & X-rays with\, \chandra{} & 554 \& 745 &
$L_{0.3-10\,\rm keV}= (15-2)\times 10^{39}$ & $2.7\times10^{-4}$ & 
\cite{Bochenek+18}\\
2020eyj & Radio detected, with e-MERLIN& 605 \& 741 & $L_{5\,\rm GHz}\approx 1.5\times10^{27}$ & $10^{-3}-10^{-2}$ & \cite{kool+2023}\\
\hline
\end{tabular}
\caption*{\textit{Notes: X-ray luminosity is in the units of\, \ergs{} and the radio spectral luminosity is in the unit of\,\ergscmhz{}.
Mass-loss rate is listed for 10\,\kms{} wind speed. } }
\end{table*}

SNe Ia have historically eluded detection in radio and X-ray wavelengths. 
 A notable case  is  SN 2014J  in M82,  the closest SN Ia in centuries, which was discovered within hours of its explosion, enabling extensive multi-wavelength follow-up observations at both early and late times, including deep radio and X-ray searches \cite{margutti+14}.  SN 2014J remained undetected in these bands \cite{margutti+14}.

Towards statistically significant searches, Chomiuk et al. \cite{Chomiuk+16} carried out Karl G. Jansky
Very Large Array (VLA) survey of 85 young ($\le1$ yr old) SNe Ia and constrained the mass-loss rate to  $\dot M/v_{\rm wind}\le 5 \times 10^{-9}$\,\ml{} (for a wind speed of 100 \,\kms{}).
Even though their sample  had 6 SNe Ia-CSM (nearest in the sample being SN 2008J at 65 Mpc), none showed detectable radio emission, which they 
attributed  to the rarity of this class. 
Lundqvist et al. \cite{Lundqvist+20} also carried out a deep radio survey of 
five nearby SNe Ia with the Multi-Element Radio
Linked Interferometer Network (MERLIN) and the Australia Telescope Compact Array
(ATCA) and constrained  mass-loss rate limits ranging $\dot M/v_{\rm wind}\le 10^{-8} -10^{-7}$\,\ml{} (for 100\,\kms{} wind speed). 
 Most recently, Griffith et al.  \cite{Griffith2025} carried out a radio study of 29 archival Ia-CSM SNe which resulted in no detection, constraining $\dot M/v_{\rm wind}\le 10^{-4} -10^{-2}$\,$M_\odot\,\rm yr^{-1}$.

 SN 2005ke was the first normal SNe Ia from which tentative X-ray emission was reported, though no radio emission was seen \cite{Immler+2006}. Its X-ray detection remains unconfirmed. SN 2012ca, which was initially classified as SNe IIn and then reclassified as Ia-CSM,  remains the only Ia-CSM with detected X-ray emission, which was seen with\, \chandra{} during $500--800$ days post discovery \cite{Bochenek+18}. The radio detection from the SN hasn't been reported so far. 
The Ia-CSM SN 2020eyj is the first and the only SN Ia of any subtype  to have shown radio emission \cite{kool+2023}. The radio emission was detected on  605 and 741 days with 5 GHz radio spectral luminosity in the range $(1.2-1.6)\times10^{27}$\,\ergshz{}. The SN was not detected in X-ray bands, though the limit was not very constraining ($L_{\rm 0.3-10\,keV}< 2.4\times10^{41}$\,\ergs). In table \ref{tab:radio_xray_IaCSM}, we tabulate the properties of radio and/or X-ray detected  SNe Ia. We also include the tentative X-ray bright SN 2005ke.

Owing to high densities, SNe Ia-CSM are candidates for efficient dust production \cite{Fox2015}. 
A rise in the mid-IR accompanied by an accelerated decline in the optical radiation of the SN  2018evt has provided clear signatures of dust formation in this SN, most likely due to ejecta-CSM interaction, three years post explosion \cite{Wang2024_2018evt}.

It is interesting to note that very few SNe Ia-CSM have shown radio  emission, despite indicating the presence of  dense CSM. One reason could be excessive absorption, which can lead to non-detectable radio emission due to progressively decreasing synchrotron strength with time
\cite{Griffith2025}. Another possibility is different microphysics of shocks in thermonuclear SNe than that in CCSNe, which may alter the efficiency of  synchrotron production. 
Finally, due to binarity playing a decisive role in SNe Ia, in Ia-CSM, the CSM may be quite complex, and geometric effects could be a major cause for non-detection.
A larger sample  spanning wider radio frequencies is required to increase the chances of detection in SNe Ia-CSM \cite{Griffith2025}.


\subsection{H-rich (Type IIP/IIL/II) supernovae}

H-rich SNe IIP, IIL, and II    typically originate from RSGs \cite{Smartt2009} and show CS interaction in various ways.  Early observations of many H-rich SNe reveal flash spectra
consisting of narrow emission lines, such as H II, He II, C III/N III, etc., within hours to days   \cite{Khazov2016}.  These lines are signatures of ionized CSM, which disappear within 2--10 days, indicating the presence of confined CSM in the vicinity of the star. Flash spectra are observed in approximately 15\% of H-rich SNe, indicating they are more ubiquitous than previously thought
 \cite{Khazov2016, Yaron2017_flash, JacobsonGalan2022}.

Amongst recent examples, flash ionization has been  studied extensively in SNe IIP  SN 2023ixf and SN 2024ggi. These are the closest SNe of the decade. SN 2023ixf was initially classified as type II \cite{Perley2023TNS}. Li et al. \cite{Li2025} presented multi-band photometry and spectroscopy out to 600 d after explosion and concluded that SN 2023ixf was a transitional SN that bridges the gap between SNe IIP and SNe IIL, owing to its comparatively short ($\sim70$ d) plateau and faster spectroscopic evolution. 
However, Kozyreva et al. \cite{Kozyreva2025} modeled the light-curve and CSM interaction and  referred to SN 2023ixf as a normal Type IIP SN.
SN 2024ggi was initially classified as a Type II SN  \cite{Tonry2024, Zhai2024}. Multi-band light curve observations at a later epoch led to its reclassification as a standard IIP SN with a prolonged recombination plateau \cite{Pessi2024, Chen2024, JacobsonGalan2024}. 
First flash signatures in SN 2023ixf were seen during 0.4--0.9\,d  arising from the ionization of  dense, confined CSM within $<10^{15}$\,cm, which completely disappeared by day 8 and unraveling the dominance of  classical Type IIP P-Cygni features \cite{JacobsonGalan2023, Bostroem2023}. The flash spectra indicated a mass-loss rate in the range of 
$ 10^{-3}-10^{-2}$\,\ml{} in the final moments of the exploding star. The high-resolution data showed velocity shifts and unequal line widths,  indicative of an aspherical CSM \cite{Smith2023}. 
SN 2024ggi also revealed flash ionization features, with a  compact confined CSM within $<5\times10^{14}$ cm \cite{Shrestha+24, JacobsonGalan2024}.
Modeling of flash spectra implied a mass-loss episode of the order of 
$10^{-2}$\,\ml{}
within $\sim 3$ years before the core-collapse \cite{JacobsonGalan2024,Pessi2024}.
The flash phase of SN 2023ixf was twice as long as that of SN 2024ggi, indicating a more confined CSM in SN 2024ggi. 
The estimated mass-loss rates in both SNe indicated an eruptive mass-loss rate for a few years leading to the explosion, which was  above a steady RSG wind.
While SNe 2023ixf and 2024ggi   are the two recent well-studied cases of flash spectra, such studies have been carried out for a significant sample. 
In table \ref{tab:FI_SNe}, we show a subset of well-studied H-rich SNe showing flash-ionization signatures.

\begin{table}
\small
\centering
\caption{A subset of Hydrogen–rich CCSNe  “flash-ionization” (FI)
signatures}
\label{tab:FI_SNe}
\begin{tabular}{lccp{5.5cm}l}
\hline
SN & Sub-type &  First FI & Representative FI lines & Ref.\\
\hline
SN~2006bp                 & IIP            & $+2$ d   & He\,\textsc{ii} ($\lambda4200$, $\lambda4686$\AA), C\,\textsc{iv} ($\lambda5805$\AA) & \cite{Quimby2007}\\
SN~2013fs     & IIP            & $+0.3$ d & He\,\textsc{ii} ($\lambda4686$\AA), N\,\textsc{v} ($\lambda\lambda4604,\,4620$\AA), N\,\textsc{iii} \& C\,\textsc{iii} ($\lambda\lambda4634,\,4640$\AA/$\lambda\lambda4647,\,4650$\AA), O\,\textsc{iv/v/vi}         & \cite{Yaron2017_flash}\\
SN~2014G                  & IIL            & 2--3 d   & He\,\textsc{ii} ($\lambda4686$)\AA, C\,\textsc{iv}\,$\lambda5803$, N\,\textsc{iii, iv} ($\lambda\lambda4640$\AA, $\lambda\lambda4057, 5201, 7113$\AA) & \cite{Terreran2016}\\
SN~2016bkv                & low-lum.\ IIP  & $\le5$ d & H\,($\lambda$4101, $\lambda4340$, $\lambda$4861, $\lambda$6563\AA), He\,\textsc{ii} ($\lambda4686, \lambda5411$\AA), C\,\textsc{iii}/N\,\textsc{iii} blend & \cite{Hosseinzadeh2018_SN2016bkv}\\
SN~2017ahn       & fast-decl.\ II & $+1$ d   & Narrow Balmer series, He\,\textsc{ii}\,$\lambda4686$                         & \cite{Tartaglia2021}\\
SN~2020pni                & II             & $+1$ d   & H$\alpha$, He\,\textsc{ii}, C\,\textsc{iii}/N\,\textsc{iii} blend            & \cite{Terreran2022}\\
SN~2020tlf                & II-P/L         & $+0.7$ d & Narrow symmetric H\,\textsc{i} and He\,\textsc{ii} lines                     & \cite{JacobsonGalan2022}\\
SN~2022acko               & low-lum.\ IIP  & $+2$ d   & He\,\textsc{ii}\,$\lambda4686$, C\,\textsc{iii}                              & \cite{Lin2025}\\
SN~2022jox                & II             & $+0.8$ d & H\,\textsc{i}, He\,\textsc{ii}, C\,\textsc{iv}, N\,\textsc{iv}               & \cite{Andrews2024}\\
SN~2023ixf                & IIP    & $+0.4$ d & He\,\textsc{ii}\,$\lambda4686$, C\,\textsc{iv}\,$\lambda\lambda5801,5812$, N\,\textsc{v}\,$\lambda\lambda4603,4619$ & \cite{JacobsonGalan2023}\\
SN~2024ggi                & IIP            & $+0.8$ d & He\,\textsc{ii}\,$\lambda4686$, C\,\textsc{iv}\,$\lambda\lambda5801,5812$, N\,\textsc{iv}\,$\lambda\lambda7109,7123$, O\,\textsc{v}\,$\lambda5590$ & \cite{JacobsonGalan2024}\\
\hline
\end{tabular}
\end{table}

Another optical signature of CS interaction in SNe IIP is optical high-velocity absorption features,
 such as H$\alpha$  and He\,\textsc{i} $\lambda$10830\AA{},   during the photospheric phase. Chugai \cite{Chugai2007_HV} indicated that these lines arise from the X-ray–excited unshocked ejecta or the CD.  
Detection of these features in
SN 1999em and SN 2004dj allowed  estimation of the mass-loss rate to be  
$10^{-6}$\,\ml{} for wind velocity of 10\,\kms{} \cite{Chugai2005,Chugai2007_HV}. 

Radio and X-ray emission have been seen in a handful of SNe IIP. For example, SN 2016X was detected with the VLA during 21--75 days, leading to a mass-loss rate estimation of 
$\sim 4.4 \times 10^{-7}$\,\ml{} \cite{RuizCarmona2022_SN2016X_radio}. In type IIP SN 2011ja, the ATCA  radio modeling revealed non-steady CSM with variable mass-loss rate in the range  $\sim  (10^{-7}-10^{-5}$\,\ml{} \cite{Chakraborti2013}. 
\chandra{} observations in SN 1999em revealed mass-loss rate of 
$\sim  10^{-6}$\,\ml{}
\cite{Pooley2002_SN1999em}.
 In Type IIP SN 2004dj, Nayana et al. \cite{Nayana2018} reported a mass-loss rate of $10^{-6}$\,\ml, which was 3 times higher than the estimate obtained by Chakraborti et al. \cite{Chakraborti2012} using the X-ray data. This  suggested a possible asymmetry, although uncertainties in various parameters and certain assumptions can easily account for this kind of discrepancy.
Strong CS interaction signatures have also been seen in the X-ray and radio bands in SNe IIL. 
SNe IIL seem to show a higher mass-loss rate than SNe IIP.
SN 1979C and SN 1980K  were the earliest examples of the best studied  SNe in this category.
 SN 1979C  had a mass-loss rate of $\sim10^{-4}$\,\ml{} derived from radio and X-ray measurements \cite{Immler2005, Weiler1986}.

H-rich  SNe typically exhibit adiabatic, or energy-conserving, interaction.
X-ray emission is usually dominated by  RS; however, inverse Compton (IC) scattering of optical photons by relativistic electrons can also contribute to X-rays at early times (and lead to steepening in radio spectrum due to loss of energy from the synchrotron emitting electrons), particularly during the optical plateau phase when there is an abundance of photons. If detected, IC, in conjunction with radio measurements, can provide insights into the distribution of energy between the magnetic field and relativistic electrons 
 \cite{Chevalier2006_IIP}.
In a typical energy  range of X-ray telescopes, 
the IC luminosity can be measured as \cite{Chevalier2006_IIP}
 \begin{equation}
L_{\mathrm{IC}}^{0.3-10\,\mathrm{keV}} \approx 3.09 \times 10^{38} \, \epsilon_r \, \gamma_{\mathrm{min}} 
\left( \frac{\dot{M}_{-5}}{v_{\mathrm{wind,1}}} \right) V_{s4} 
\left( \frac{L_{\mathrm{bol}}(t)}{10^{42}~\mathrm{erg\,s^{-1}}} \right)
\left( \frac{t}{10~\mathrm{days}} \right)^{-1}
~\mathrm{erg\,s^{-1}}.
\end{equation}

Signatures of cooling, when cooling timescales are shorter than the dynamic/expansion timescale, have also been seen in radio observations of SNe II/IIP/IIL.  The ratio of the Compton cooling timescale and expansion timescale is \cite{Chevalier2006_IIP}
\begin{align}
\frac{t_{\mathrm{Comp}}}{t} \approx 0.1 
&\left( \frac{L_{\mathrm{bol}}}{2 \times 10^{42}~\mathrm{erg\,s^{-1}}} \right)^{-1} \notag \\
&\times \left( \frac{\epsilon_B}{0.1} \right)^{1/4}
\left( \frac{\dot{M}_{-5}}{v_{\mathrm{wind,1}}} \right)^{1/4} 
 \left( \frac{V_{\rm sh}}{10^4\,\rm km\,s^{-1}}\right)^2
\left( \frac{\nu}{10\,\mathrm{GHz}} \right)^{-1/2}
\left( \frac{t}{10\,\mathrm{days}} \right)^{1/2}
\label{eq:tcomp}
\end{align}
Here $L_{\mathrm{bol}}$ is the bolometric luminosity and $\epsilon_B$ is the fraction of magnetic field energy density of the thermal energy. 

Synchrotron cooling can also be important in some cases, which can be determined from the ratio of the synchrotron timescale to the expansion timescale,  given by \cite{Chevalier2006_IIP}
\begin{equation}
\frac{t_{\mathrm{synch}}}{t} \approx 11.3 \left( \frac{\epsilon_B}{0.1} \right)^{-3/4} 
\left( \frac{\dot{M}_{-5}}{v_{\mathrm{wind,1}}} \right)^{-3/4} 
\left( \frac{\nu}{10\,\mathrm{GHz}} \right)^{-1/2} 
\left( \frac{t}{10\,\mathrm{days}} \right)^{1/2}.
\end{equation}

Since
$t_{\mathrm{Comp}} \propto \epsilon_B^{1/4}$
and
$t_{\mathrm{synch}} \propto \epsilon_B^{-3/4}$, higher energy in the magnetic field will lead to dominant synchrotron cooling and higher energy in relativistic electrons to Compton cooling.  

Cooling steepens the particle energy index $p$ by 1. This is observable in radio spectra via steepening of the optically thin spectral index $\alpha$ ($\alpha=(p-1)/2$) by 0.5, a phenomenon observed in several SNe II/IIP/IIL.
Nayana et al. \cite{Nayana2018} presented long-term radio light curves of SN 2004dj and  found signatures of cooling due to IC in the early time of radio data. 
Chakraborti et al. \cite{Chakraborti2012} analyzed the archival X-ray data of a Type IIP  SN 2004dj and found that the  X-ray emission originated from a combination of IC and thermal emission from the RS. They found a progressively decreasing contribution of the IC component with time. By modeling IC X-ray along with radio measurements, they were able to determine the fractions of energy into relativistic electrons $\epsilon_e$ and the magnetic field $\epsilon_B$ to be 0.082 and 0.39, respectively.
However,  cooling signatures  seen in SN 1979C for over 10 years were synchrotron in nature, probably indicating different microphysics than that of SN 2004dj, with a larger fraction of energy in 
$\epsilon_B$ than in $\epsilon_e$ \cite{Weiler1991}.  In SN 1979C, the X-ray emission was primarily thermal, with a less significant hard X-ray component, and no strong IC component was indicated  \cite{Immler2005}.  While in SN 1999em, SN 2013ej and 2020fqv,
X-ray measurements  also  revealed a soft thermal bremsstrahlung along with a harder IC tail
 dominating during optical luminosity peak \cite{Pooley2002_SN1999em, Chakraborti2016, JacobsonGalan2022}, in SN 2004et, X-ray emission could be modelled with purely thermal plasma \cite{Misra2007}.

In H-rich SNe, absorption of radio emission has revealed FFA to be a dominant mechanism, though observable signatures of SSA have been in some cases,  In SN 1979C, radio observations indicated external FFA from a dense $\rho \propto r^{-2}$ \cite{Weiler1981}. In SN 1980K, radio data also favored  FFA with a wind density comparable to SN 1979C  \cite{Weiler1986}. 
 In another well-sampled H-rich SN 2004et, in which radio data favored a combined FFA$+$SSA model, along with  the onset of synchrotron cooling at $\ge 10$ days \cite{MartiVidal2007}.
  In SN 2020fqv, the radio peak (on day 35) was found to be dominated by FFA from a confined CSM shell \cite{JacobsonGalan2022}.

SN 2023ixf and SN 2024ggi are two H-rich SNe that have provided valuable insights into the nature of the progenitor via high cadence multiwavelength observations.
In SN 2023ixf, \,\nustar{} detected hard X-rays on days 4 and 11, the earliest such detection for a SN of any type \cite{Grefenstette2023}. The X-ray emission suggested a shock temperature  $\gtrsim 25$\,keV and a column density  $N_H=(2-3)\times10^{23}$\,cm$^{-2}$, consistent with high density CSM and  FS dominated early emission.
Further observations with\, \chandra{} on days 13 and 86 showed a decrease in column density following $t^{-1}$, resulting in a constant, albeit smaller, mass-loss rate ($\dot M=6.5\times 10^{-4}$\,\ml) \cite{Chandra2024}.
Fe K$\alpha$ 6.4\,keV line was observed during both \,\nustar{}  epochs and the first \, \chandra{} epoch, which was interpreted as  the CSM  not being fully ionized within the first 13 days. This phenomenon was also seen in SN 2010jl, a type IIn SN, though at a much later stage \cite{Chandra2015}. Nayana et al. \cite{Nayana2024} presented comprehensive X-ray data between day 4 to 165 using \,\nustar{}, \, \chandra{},\,\xmm{},
and\,\swift. The analysis showed that the broadband X-ray spectra remained thermal throughout and  both column density and temperature  declined with time.
The peak X-ray luminosity was measured to be $3\times10^{40}\,$\ergs{}, making it the most luminous H-rich SN ever recorded. Nayana et al. \cite{Nayana2024} also compiled radio data covering the frequency range $0.6-84$\,GHz. The SN became radio detectable around day $15-20$, and it revealed synchrotron emission with a time-dependent FFA. The data were consistent with the multi-zone CSM and/or inhomogeneous ejecta.

In SN 2024ggi, the first X-ray emission was reported by the Astronomical Roentgen Telescope – X-ray Concentrator  in $4-12$\,keV range during $1.5-2.5$ days \cite{Lutovinov2024}. The \ep{} and the \nustar{}  revealed bright X-ray emission along with Fe\,K$\alpha$ 6.4\,keV line \cite{Zhai2024}. In radio bands, after initial  non-detection from mm to cm bands \cite{Hu2024, Chandra2024}, Ryder et al. \cite{Ryder2024} reported the first detection with the ATCA around day 25. Both SNe exhibited early, highly absorbed hard X-ray emission originating from the FS. Notably, SN 2023ixf was almost an order of magnitude brighter than SN 2024ggi in both radio and X-ray wavelengths \cite{Lutovinov2024, Zhang2024, Chandra2024, Nayana2024}.

In Table \ref{tab:II_SNe}, we list some well-studied hydrogen-rich SNe and their radio and X-ray properties. We also indicate the cases where X-ray emission showed an IC component and/or radio observations revealed cooling.

\begin{table}[ht]
\centering
\small
\caption{List of well studied H-rich SNe with radio and X-ray emission}
\begin{tabular}{lllllcllcl}
\toprule
& & & \multicolumn{3}{c}{\textbf{Radio}} & 
\multicolumn{3}{c}{\textbf{X-ray}} & \\
 \cmidrule(lr){4-6}
  \cmidrule(lr){7-9}
\textbf{SN} & \textbf{Type} & \textbf{Distance} & \textbf{$L_{\nu,\mathrm{p}}^R$} & \textbf{$\Delta t_{\rm p,R}$} & Cooling & \textbf{$L_{\mathrm{p}}^{X}$} & \textbf{$\Delta t_{\rm p,X}$}&  
\textbf{IC} & \textbf{Ref.} \\
 & & Mpc & erg\,s$^{-1}$\,Hz$^{-1}$ & days & & 
 erg\,s$^{-1}$ & days & & \\
\hline
1979C & IIL & 17.1 & $2.8\times10^{27}$ & 450 & -- & $0.8\times 10^{39}$ & 5900 & -- & \cite{Montes2000, Immler2005}\\
1980K & IIL & 5.1 & $1.0\times10^{26}$ & 130 & -- & $3.5\times10^{39}$ & 44 & -- & 
\cite{Schlegel1995, Weiler1992, Weiler1986}\\
1999em & IIP & 11.7 &  $2.2\times10^{25}$  & 34 & Yes & $9\times10^{37}$ & 4 & Yes &\cite{Pooley2002_SN1999em}\\
2002hh & IIP & 5.5 & $1.5\times10^{25}$ & 29 & -- &
$4\times10^{38}$ & 25 & -- & \cite{Bietenholz2021, Chevalier2006_IIP}\\
2004dj& IIP &  3.1 & $2.1\times10^{25}$ & 37 & Yes &
$1.5\times10^{38}$ & 42 & Yes 
 & \cite{Chakraborti2012, Nayana2018}\\
2004et& IIP & 5.5 & $8.7\times10^{25}$ & 45 & -- &
$3\times10^{38}$ & 30 & -- & \cite{Misra2007} \\
2011ja & IIP & 3.4 &   $1.1\times10^{25}$ & 30 & Yes
& $5.5\times10^{38}$ & 113 & Yes & \cite{Chakraborti2013}\\
2012aw & IIP & 10 &  $7.2\times10^{25}$ & 23 & Yes & 
$9.2\times10^{38}$ &  5 & Yes & \cite{Yadav2014, Immler2012}\\
2013ej & IIP/IIL  & 9.6 & $5\times10^{25}$ & 27 & -- &  $7.8\times10^{38}$ & 13 & Yes & \cite{Chakraborti2016, Bietenholz2021}\\
2016X & IIP & 15.2 & $2.1\times10^{26}$ & 27.6 & Yes  & $2.5\times10^{39}$ &$2-5$ & Yes & \cite{RuizCarmona2022_SN2016X_radio}\\
2017eaw & IIP & 5.9 & $1.7 \times10^{25}$ & 20 & -- & $1.1\times10^{39}$ & 2 & -- & \cite{Szalai2019}\\
2023ixf & IIP & 6.9 & $1.1\times10^{26}$ & 165 & Yes & $1\times10^{40}$ & 11.5 & -- & \cite{Chandra2024, Nayana2024}\\
2024ggi & IIP & 7 & $6\times10^{24}$ & 23 & - & -- & 4 & - & \cite{Zhang2024, Ryder2024}\\
\bottomrule
\end{tabular}
\caption*{\textit{Notes. Radio peak spectral luminosity is mentioned for 8 GHz frequency and X-ray peak luminosity is in 0-3--8\,keV range. } }
\label{tab:II_SNe}
\end{table}

\subsubsection{SN 1987A}
While not a typical IIP, IIL, or II type, with a classification of IIP-pec, SN 1987A  deserves special mention, being the “celebrity” SN  in the Large Magellanic Cloud due to its proximity, timing, multi-messenger signals, and astrophysical surprises. It is the closest CCSN since Kepler, which allowed high-S/N spectroscopy and imaging from $\gamma$-rays to radio bands.
Hours before optical discovery, a burst of  a couple of dozen neutrinos was detected  \cite{Hirata1987}, confirming the most basic paradigm of the core-collapse mechanism. A BSG, Sk-69$^o$ 202,  was identified as the SN progenitor \cite{Walborn1987}.
Due to unprecedented multiwavelength coverage, several signatures of CS interaction were seen, including UV flash, early SSA radio burst, decades-long radio/X-ray rise, and dust IR echo \cite{McCray2016,Turtle1987,Park2005}. The SN shows a remarkable triple-ring nebula, featuring an equatorial ring and two polar rings.  Observations  in multiple wavelengths revealed that the   expanding FS is interacting with  the equatorial ring, creating the famous hotspots and strong X-ray/optical emission from shock-heated gas \cite{Fransson2015}.
 Recently,  a  “direct” evidence for a neutron star in SN 1987A was seen using the James Webb Space Telescope, which detected a compact source at the center via  highly ionized argon and sulfur lines, formed due to  high-energy radiation from a newborn neutron star \cite{Fransson2024}.

\subsection{Stripped envelope (IIb/Ib/Ic) supernovae}

In SESNe (SNe IIb/Ib/Ic), SSA is usually the dominant absorption mechanism of the radio emission.  In SNe IIb,  FFA  can have an additional  contribution \cite{Fransson1998}.
When SSA is dominant, electrons producing synchrotron emission also participate in the absorption,  thus one can obtain the size  and the  magnetic flux by measuring
 SSA peak  density when the optical depth is 1.  The  peak radius $R_p$ and peak magnetic field, $B_p$, based on measured peak flux density, $F_{\rm p}$, at a frequency $\nu$, assuming $f$  as the filling-factor (the fraction of a given volume occupied by the  plasma), can be defined as   \cite{Chevalier2006_Ib}. 
\begin{equation}
R_p = 4.0 \times 10^{14} \, \left(\frac{\epsilon_{\rm e}}{\epsilon_B}\right)^{-1/19} \left( \frac{f}{0.5} \right)^{-1/19} \left( \frac{F_{\mathrm{p}}}{\mathrm{mJy}} \right)^{9/19} \left( \frac{D}{\mathrm{Mpc}} \right)^{18/19} \left( \frac{\nu}{5\,\mathrm{GHz}} \right)^{-1} \, \mathrm{cm}
\end{equation}
and 
\begin{equation}
B_p = 1.1 \, \left(\frac{\epsilon_{\rm e}}{\epsilon_B}\right)^{-4/19} \left( \frac{f}{0.5} \right)^{-4/19} \left( \frac{F_{\mathrm{p}}}{\mathrm{mJy}} \right)^{-2/19} \left( \frac{D}{\mathrm{Mpc}} \right)^{-4/19} \left( \frac{\nu}{5\,\mathrm{GHz}} \right) \, \mathrm{G}
\end{equation}

X-ray observations  of SESNe usually require  a non-thermal mechanism. IC can be a strong candidate for non-thermal emission at early times, though the synchrotron mechanism may dominate at late epochs \cite{Berger2002, Sutaria2003, Bjornsson2013}.   Bj\"ornsson et al.  \cite{Bjornsson2013,Bjornsson2017} have argued that  radio emission in SESNe may suffer from inhomogeneities in the synchrotron emitting shocked zone and may alter important inferences like size of the emitting region, magnetic field, etc. Inhomogeneities will 
result in a range of optical depths  broadening the observed radio spectrum, and 
also boost IC emission in X-rays, which combined with radio observations, can be used to quantify the inhomogeneities. Chandra et al. \cite{Chandra2019}  found observational evidence of inhomogeneities in Type Ib SN Master OT J120451.50+265946.6, and adapted the model of 
Bj\"ornsson \& Keshavarzi  \cite{Bjornsson2017} to interpret their radio observations. This  model assumes that the inhomogeneities can be due to variations in the distribution of magnetic fields and/or  relativistic electrons, and can be quantified with a parameter $\delta'$ representing the correlation between the distribution of relativistic electrons and the distribution of magnetic field strengths. This parameter is  such that  for  $\delta'=0$ the inhomogeneities between the two distributions are not correlated,  whereas for $\delta'=1$  the inhomogeneities between the magnetic field and the relativistic electron distributions are correlated.
In the range of magnetic fields, when inhomogeneities affect the radio spectrum, the radio flux density takes the form 
$F_\nu\propto \nu^{\frac{3p + 7 + 5\delta' - a(p + 4)}{p + 2(1 + \delta')}}$ \cite{Bjornsson2017,Chandra2019}. 
The inhomogeneous model  was later applied to other transients, such as FBOT  AT 2018cow \cite{Nayana2021} and SN 2023ixf \cite{Nayana2024} and

Amongst  SESNe, SNe IIb  contain only trace amounts of hydrogen in their spectra. They serve as an important link between H-rich SNe and other SESNe.
Chevalier \& Soderberg \cite{Chevalier2010} proposed two types of progenitors for SNe IIb, categorized by optical properties and CS interaction: compact progenitors (cIIb SNe) and extended progenitors (eIIb SNe).
SNe cIIb, e.g., SN 2008ax, SN 2001ig, SN 2003bg, and SN 2011dh, behave more like SNe Ib/c, and their progenitors are likely  either a single WR star or a stripped star in an interacting binary \cite{Chevalier2010}. 
SNe eIIb  resemble SNe II and arise from supergiants. SNe in this class are SNe 1993J and 2001gd. 
They emphasized that radio properties can distinguish between the two types; however,  subsequent works, such as Ouchi et al. \cite{Ouchi2017} etc., have suggested that the strict cIIb and eIIb distinction is overly simplistic. The studies argue that IIb progenitors occupy a continuum  in terms of radius and mass-loss properties.

\begin{figure}
\begin{center}
\includegraphics[angle=0,width=1\textwidth]{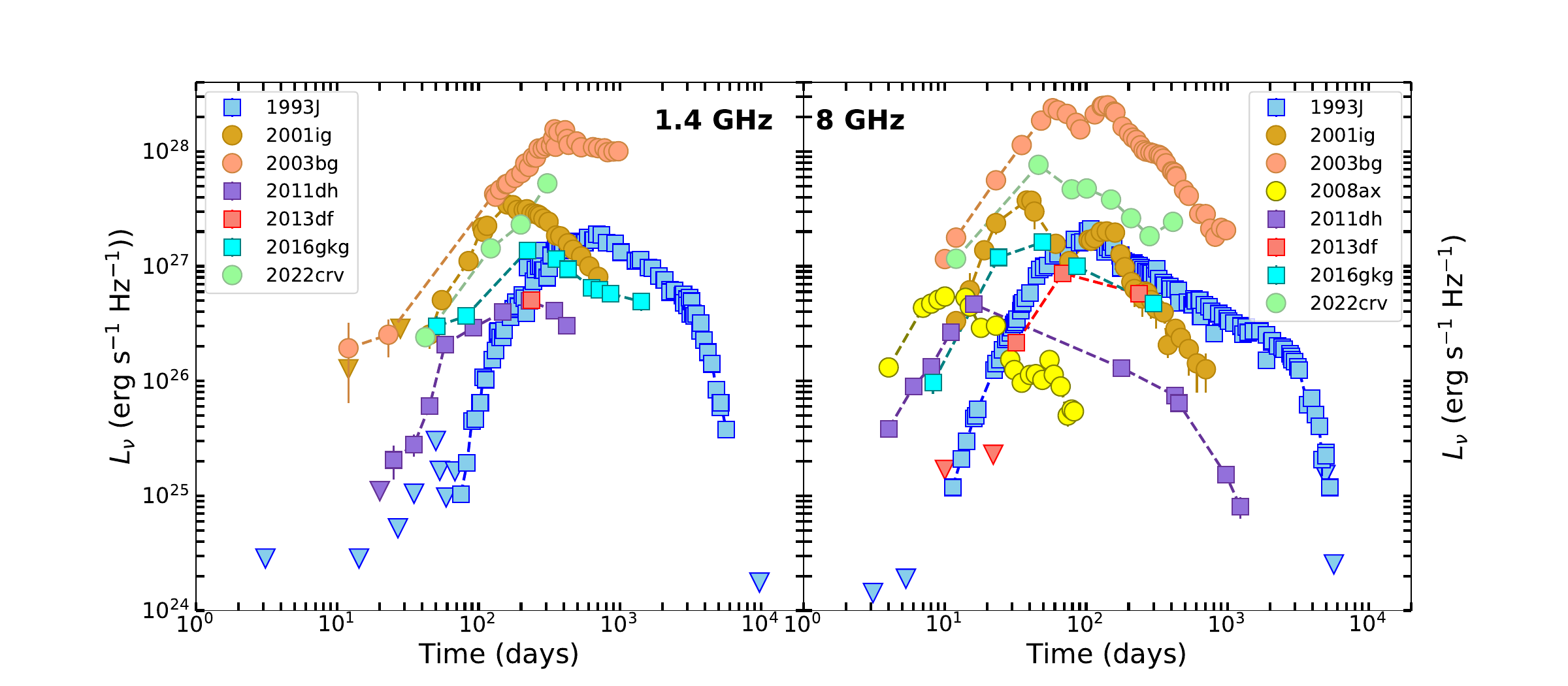}
 \caption{1.4 GHz (left) and 8 GHz (right) radio spectral luminosity of published SNe IIb.  SNe eIIb are plotted with squares, and SNe cIIb are plotted with circles. There is no  clear differentiation between compact and extended progenitors. However, there may be a hint of faster evolution of SNe cIIb. SN 2008ax, with the highest ejecta velocity, clearly peaks at much earlier times than other SNe IIb. The data are taken from \cite{Nayana2022, Gangopadhyay2023}.}
   \label{fig:NuLnu}
\end{center}
\end{figure}

Observationally, SNe cIIb show rapid evolution of CS interaction, resembling SNe Ib/c and have FS speeds $\gtrsim 30,000$\,\kms. Whereas, SNe eIIb  show speeds consistent with  $\sim 10,000$\,\kms{} \cite{Chevalier2010}. SNe cIIb also show a more dominant non-thermal component in X-ray emission, unlike SNe eIIb, which  mainly emit thermal X-rays.
A classic example of SNe eIIb is SN 1993J, which was very well observed due to its proximity. The X-ray emission from the SN was mostly thermal \cite{Fransson1996, Chandra2009}. The RS stayed radiative for a few years before turning adiabatic, whereas FS remained adiabatic throughout the evolution \cite{Chandra2009}. The SN remained bright in radio bands for a very long time \cite{Chandra2004, Weiler2007}, and the radio emission was absorbed with a combination of FFA and SSA \cite{Fransson1998}.  The late-time radio emission revealed synchrotron cooling, indicating $\epsilon_B$ dominated over $\epsilon_e$  by at least a factor of 10 \cite{Chandra2004}. The milliarcsec very long baseline interferometry (VLBI) measurements showed FS moving with $\lesssim 10,000$\,\kms{}, and a rather symmetric  radio emitting shell \cite{Bietenholz2003}. On the contrary, the classic cIIb  SN 2008ax evolved very fast with  shock speeds  $50,000$\,\kms{}
\cite{Roming2009}. In SN 2008ax, the X-ray luminosity also declined by  a factor of 4 within a month \cite{Roming2009}, revealing 
either a smaller extent of CSM or much faster winds consistent with those of WR stars.
Other well-studied examples of SNe IIb are SN 2016gkg \cite{Nayana2022}, where radio evolution and high shock speeds indicated a compact progenitor, while a detailed study of another well-studied SN 2013df suggested  an extended progenitor \cite{Kamble2016}.
A very well observed SN 2011dh  proved to be a link between SNe  cIIb and eIIb  \cite{Horesh2013,Soderberg2012}.
 SN 2001ig deserves special mention among SNe IIb as it showed modulations with a period of 150 days \cite{Ryder2004}.  Soria et al. \cite{Soria2025} found radio rebrightening $>20$\,yrs post explosion. This behavior has been best explained in a binary scenario in which
the SN ejecta  hit a denser CSM shell, perhaps compressed by the fast wind of the WR progenitor or expelled centuries before
the SN. 
 In Fig. \ref{fig:NuLnu}, we compile the radio light curves of various SNe IIb from the literature and plot their 1.4 GHz and 8 GHz radio luminosities. While there is no clear differentiation between the two classes, there is a hint that SNe cIIb   have faster evolution with radio peaks reaching earlier times.

Compared to SNe IIb, SNe Ib/c show  relatively faster, low-density winds. Their radio emission  usually peaks earlier than SNe IIb, and the absorption is dominated by the SSA.  
SNe Ib/c gained significant attention after the association of GRB 980425 with SN 1998bw \cite{galama1998sn1998bw} and GRB 030329 with SN 2003dh \cite{Stanek2003,Hjorth2003}.
Berger et al. \cite{Berger2003} and Soderberg et al. \cite{Soderberg2006} carried out   late-time radio observations of several dozen SNe  Ib/c, in order to search for off-axis GRBs whose jets were spreading into the  line of sight. Based on their non-detections in the majority of the cases, they  concluded that 
$\lesssim 10$\% of  these SNe Ib/c are associated with off-axis GRBs. 

Individual  SNe Ib/c with detailed radio follow-up have shown SSA dominance, and the derived shock speeds are close to $
\sim 0.1-0.3$\,c, though the mass-loss rates and luminosities are not very different than H-rich SNe. SN 2012au is a well-studied example of this class, in which 
Kamble et al. \cite{Kamble2014} presented analysis and the SN  revealed X-ray luminosity  $10^{39}$\,\ergs{} and  a bright radio emission. They   found that the energy of the radio-emitting material (moving with $0.2$\,c)  is intermediate between normal CCSNe and SNe, which are associated with GRBs (GRB-SNe). 

Amongst 
SNe Ic,  SN 1994I, is one of the oldest examples.  The SN distance ranges between 7.8 and 8.4 Mpc, based on its host galaxy M51, and the Expanding Photospheres Method \cite{VanDyk2016}, making it one of the nearby SN of this class.   SN 1994I X-ray studies were presented by Schlegel et al. \cite{Schlegel1999} and Immler et al. \cite{Immler2002}.  
Weiler et al. \cite{Weiler2011} presented the radio studies covering 3000 days of observations.  The SN peaked within a couple of weeks of discovery itself, and it remained bright for a long.
Other older SESNe worth mentioning, for which well-sampled radio light curves exist, are  
SN 1983N \cite{Weiler1986}, SN 1984L \cite{Panagia1986}, SN 1990B \cite{vanDyk1993} etc.

Among more recent ones, Berger et al. \cite{Berger2002}  investigated the radio observations of SN 2002ap, a Type Ic SN,  and  found the shock speed to be 0.3\,c on day 2.
Sutaria et al. \cite{Sutaria2003}  discovered the SN with the \xmm{} on day 4 with a luminosity of  $6.4 \times 10^{37}$ erg\,s$^{-1}$  and found the 
The X-ray spectrum had a significant contribution from a non-thermal IC component. 
 In Ic  SN 2020oi, while no  X-ray emission was seen, a bright radio emission emerged from the SN
\cite{Horesh2020}. From the measurement of IC cooling, they  showed  a deviation from equipartition  in the microscopic parameters.
In this SN, the early time  ALMA observations revealed 
confined CSM within $10^{15}$ cm \cite{Maeda2021}. As the progenitors of SNe Ic are bare C$+$O stars, this translates into the enhanced mass-loss activity in the last year. 
Such studies have important implications for understanding SNe and their associations with the central progenitor engines.

As mentioned above, multiple surveys were carried out for SESNe to look for an off-axis GRB jet. However, it has  
become  clear that GRB-SN are mostly Ic-bl type, 
though, there are  exceptions  like   GRB 970514 associated with Type IIn SN 1997cy \cite{Germany2000, Turatto2000}, GRB 980910 associated with peculiar Type II SN 1999E \cite{Rigon2003}, and GRB 111209A, associated  with  SLSN I SN 2011kl (which was likely powered by magnetar) \cite{Wang2017}.  Consequently, the focus has been mostly narrowed down to SNe Ib-bl to search for the 
 off-axis GRBs.  
 Soderberg  et al. \cite{Soderberg2006} sample had a handful of SNe-bl, but that search  resulted in a null result, and they ruled out the scenario in which every SNe Ic-bl harbored a GRB at the 84\% confidence level. 
 Corsi \cite{Corsi2016}  carried out radio observations of  15 SNe Ic-bl discovered by PTF and 
detected only 3. The derived shock speeds  revealed that 
while GRB-SNe are usually relativistic, this is not common for SNe Ic-bl without GRBs.
Later Corsi et al. \cite{Corsi2023}
carried out a radio and X-ray  follow-up campaign of 16 SNe Ic-bl detected by the ZTF  and detected four SNe, though none of them revealed relativistic explosions like  SN 1998bw. Based on their results, they concluded that SNe Ic-bl, which are similar to GRB-associated SNe or relativistic SNe without GRB association, are rare.
Srinivasaragavan et al. \cite{Srinivasaragavan2024} carried out the largest systematic study of 36 SNe Ic-bl with the  ZTF. 
Thirteen SNe Ic-bl  in their sample had radio observations  with eight detections and five non-detections. They found that the optical properties of radio-detected vs non-detected SNe Ic-bl are indistinguishable and that only the radio data can reveal significant differences, likely connected with a central engine.
It has been suggested by these studies that  the viewing angle effects alone cannot account for GRBs associated with SNe Ic-bl and there are likely  intrinsic differences between the
 explosion mechanisms
of GRB-associated SNe Ic-bl and non-GRB SNe Ic-bl. 

Individual studies of some SNe Ic-bl have helped us delve into their CS interaction.
Salas et al. \cite{Salas2013} presented radio light curve and X-ray observations of SN Ic-bl, SN 2007bg, and found 
two distinct mass-loss episodes, with mass-loss rates ranging from $10^{-4}$\,\ml{} to  $10^{-6}$\,\ml{}
for 1000\,\kms{} wind speed.
However, unlike other SNe of this class, it  peaked very late on day 567 with bright radio luminosity $10^{29}$\,\ergshz{}, possibly due to 
additional CS interaction from a dense shell due to binarity.
Nayana et al. \cite{Nayana2020} presented GMRT and VLA observations of Ic-bl SN   ASASSN-16fp (SN 2016coi) and  carried out a GHz to sub-GHz study of the SN and presented radio light curves extending to late time. 
Detailed modeling of the data  revealed varying  shock velocities, which deviate from simplistic models. 
Corsi  et al. \cite{Corsi2014} carried out detailed radio and X-ray studies of unique Ic-bl SN PTF11qcj. They found that the SN reached radio luminosity comparable to that of SN 1998bw and was detected with the\, \chandra{} telescope in X-ray bands \cite{Kouveliotou2004}, which is compatible with extrapolation of radio synchrotron emission. The ejecta was moving with 0.3--0.5\,c. This combined with high inferred 
mass-loss rate of $10^{-4}$\,M$_\odot$\,yr$^{-1}$ makes it a very unique SN. 
Another SN Ic-bl worth mentioning is SN 2020bvc at a distance of 114 Mpc \cite{Ho2020}.
It was detected in radio as well as X-ray bands. Ho et al. \cite{Ho2020} presented $13-43$ day light curves of the SN and found mildly relativistic ejecta speeds. 
The peak radio luminosity at 10 GHz was $\sim10^{27}$\,\ergshz{}, much less than interaction-powered SNe. In contrast, X-ray luminosity was quite large 
$\sim10^{41}$\,erg\,s$^{-1}$. They suggested that non-thermal IC or synchrotron can  explain only a fraction of this  high luminosity, and the lion's share of radiation arises from thermal hot shocked plasma.
Due to their fast evolution, only a handful of non-GRB SNe Ic-bl have been observed in X-ray bands. 
Only a handful of SNe Ic-bl  have shown X-ray emission, e.g., SN 1998bw  \cite{Kouveliotou2004} (GRB-SN), SN 2020bvc \cite{Ho2020}. 
The recent discovery of EP250108a/SN 2025kg  has opened a new and very efficient channel to discover more SNe of this class \cite{Srinivasaragavan2025}.

\begin{figure}
\begin{center}
\includegraphics[angle=0,width=0.99\textwidth]{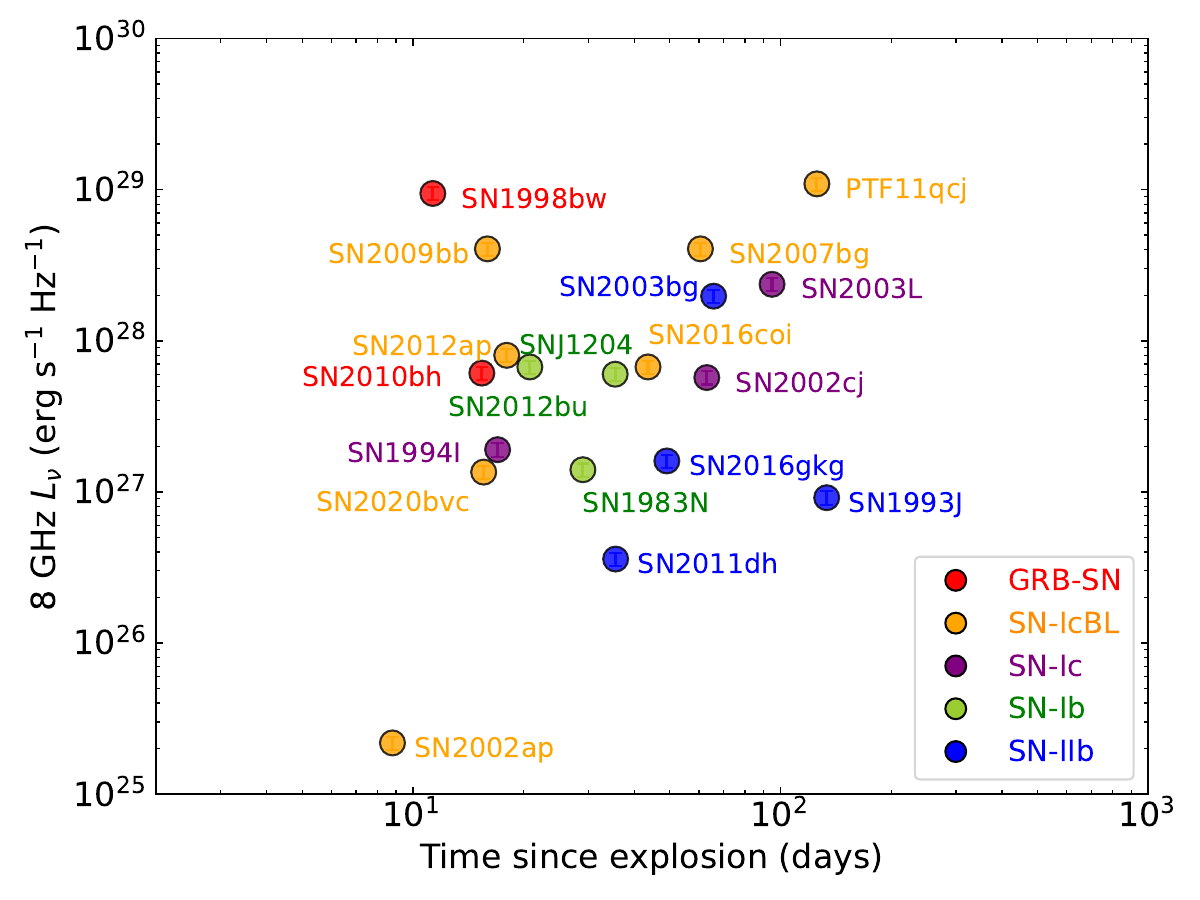}
 \caption{Peak 8 GHz radio spectral luminosity and the peak time for SESNe for a small sample of well-observed cases. While SNe IIb/Ib/Ic do not show a major difference from SNe Ic-bl, SNe associated with GRBs  seem to be intrinsically different from the rest of the sample. Most of the data are taken from  \cite{Nayana2021, Nayana2022, Ho2020}.}
   \label{fig:SESNe}
\end{center}
\end{figure}

In Fig. \ref{fig:SESNe}, we plot peak radio luminosity and the peak time for SESNe for a small sample of well-observed cases. While SNe IIb/Ib/Ic do not show a major difference from SNe Ic-bl, SNe associated with GRBs seem to be intrinsically different from the rest of the SESNe population, reaching peak
luminosities faster. Radio observations combined with X-ray measurements should be able to break the degeneracy in SNe Ic-bl.

Flash spectra have been seen in  H-rich SNe, and SESNe generally have not been observed to reveal it. There are  exceptions, such as SN 2013cu, which showed signatures of flash spectra, though it was a IIb SN, a bridge between H-poor and H-rich SN \cite{Galyam2014}.
However, ALMA mm observations of an Ic-bl SN   2020oi  have revealed its
confined CSM within $10^{15}$ cm \cite{Maeda2021}, which translates into 
the enhanced mass-loss activity in the last year.  The GHz-sub GHz observations 
later  revealed standard  
$10^{-5}$\,\ml{} mass-loss rate between
$10^{15}$ to $10^{16}$ cm, before seeing another enhanced mass-loss episode \cite{Nayana2020}.
It is likely that the enhanced mass-loss activity may be  driven by the accelerated nuclear burning  in the last moments  of the star \cite{Ouchi2019}, though later enhancement may be related to clumpy CSM or binarity.

\subsection{Interacting Supernovae}

We categorize interacting SNe as a class of SNe surrounded by dense CSM.   While SNe Ia-CSM are also interacting SNe, we discuss them  along with other thermonuclear SNe, based on the stellar death pathway. Here, we focus on interacting SNe within the CCSNe category. 
These are H-rich interacting SNe, SNe IIn, and H-poor interacting SNe, SNe Ibn and SNe Icn. In addition, we also include SNe that undergo metamorphosis and turn into interacting SNe.
We first discuss some common properties of interacting SNe and then discuss individual subclasses.

In interacting SNe, a CDS usually forms at the CD, a hallmark of radiative ejecta–CSM interaction. The CDS formation occurs when the post-shock cooling time $t_{\rm cool}$ is  shorter than the expansion time $t$.
The time scale for RS cooling depends heavily on density gradient, mass-loss rate, and ejecta speed, and can be expressed as \cite{Chevalier2003}:
\begin{equation}
t_{\mathrm{cool}}(\mathrm{RS}) = \frac{605}{(n - 3)(n - 4)(n - 2)^{3.34}} 
\left( \frac{V_{\mathrm{ej}}}{10^4\,\mathrm{km\,s}^{-1}} \right)^{5.34}
\left( \frac{\dot{M}_{-5}}{v_{\mathrm{wind,1}}} \right)^{-1}
\left( \frac{t}{\mathrm{days}} \right)^2 \, \mathrm{days}
\end{equation}
 The 
column density of the CDS is \cite{Chevalier2003}
\begin{align}
N_{\mathrm{CDS}} \sim 10^{21} (n - 4)
\left( \frac{\dot{M}_{-5}}{v_{\mathrm{wind,1}}} \right) 
 \left( \frac{V_{\mathrm{ej}}}{10^4\,\mathrm{km\,s}^{-1}} \right)^{-1}
\left( \frac{t}{100\,\mathrm{days}} \right)^{-1}
\, \mathrm{cm}^{-2}
\end{align}

In regular CCSNe, X-ray emission is typically seen in $\sim$\,keV range, dominated by  radiation coming from the RS owing to their higher  densities. In  interacting SNe, X-ray emission is usually seen to be dominated by the FS. This is because in these SNe
$t_{\mathrm{cool}}/t<1$, RS becomes  radiative, and the CDS absorbs most of the radiation coming out of RS, and FS comes to dominate.
Another factor of FS domination is the density dependence of adiabatic and radiative luminosity. While the radiative RS  luminosity scales with density, the
luminosity from the adiabatic FS continues to grow as density
squared, leading to observable emission predominantly from the FS \cite{Chevalier2017}. However, in some extreme cases, e.g., SN 2010jl \cite{Chandra2015},  the FS can also become radiative.
The kinetic luminosity of the  radiative shocks can be described by the following equation \citep{Chevalier2003}:
\begin{equation}
L_i= \pi R_{\rm s}^2(1/2 \rho_i V_i^3)
= c_i  \frac{(n-3)}{(n-2)}  \times \frac{1}{2} \frac{\dot M V_{\rm ej}^3}{v_{\rm wind}}
\end{equation}
Here $c_i$ is $(n-4)/2(n-2)^2$ and $(n-3)^2/(n-2)^2$ for the RS and  the FS, respectively. Thus radiative luminosity goes as $L_i \propto t^{-3/(n-2)}$ for steady
wind. However, for more general case,  $L_{\rm i} \propto  t^{-(15-6s+ns-2n)/(n-s)}$.

Formation of CDS in interacting SNe may lead to mixing of  absorbing thermal electrons in synchrotron-emitting region, which can lead to internal FFA. This configuration modifies the observed radio flux's frequency and time dependence, which takes the form:
\begin{align}
F_{\nu}(t) &= K_1 \nu^{\alpha} t^{\beta} \left( \frac{1 - \exp(-\tau_{\nu}^{\mathrm{intFFA}})}{\tau_{\nu}^{\mathrm{intFFA}}} \right) \\
\tau_{\nu}^{\mathrm{intFFA}} &= K_2 \nu^{-2.1} t^{\delta'}
\end{align}
Here $\delta'$ indicates the index of  internal FFA time evolution. 
Internal FFA results in a less severe reduction of radio flux density compared to external FFA, as a portion of the emitting region still remains visible.

Interacting SNe are efficient dust producers, CDS being the major dust-producing site as a result of ejecta-CSM interaction, though dust production in other H-rich SNe happens mainly in the ejecta \cite{Fox2011,Fox2013}.

\subsubsection{Type IIn supernovae}

SNe IIn show a common observational feature – narrow
H$\alpha$ emission atop of broad emission, without
broad absorption \cite{schlegel1990}. Powered by ongoing  interaction,  SNe IIn usually stay optically bright for months to years. SNe IIn  show evidence of pre-existing dust as well as formation of new dust
\cite{Gall2014, smith2020, Chandra2022,Shahbandeh25, Clayton25,Baer-way25}. Some examples of new dust formation are SN 1998S \cite{Mauerhan2012}, SN 2005ip \cite{smith2017}, SN 2010jl \cite{Gall2014}, SN 2017hcc \cite{smith2020} and SN 2020ywx \cite{Baer-way25}. 

\begin{figure}
\begin{center}
\includegraphics[angle=0,width=0.88\textwidth]{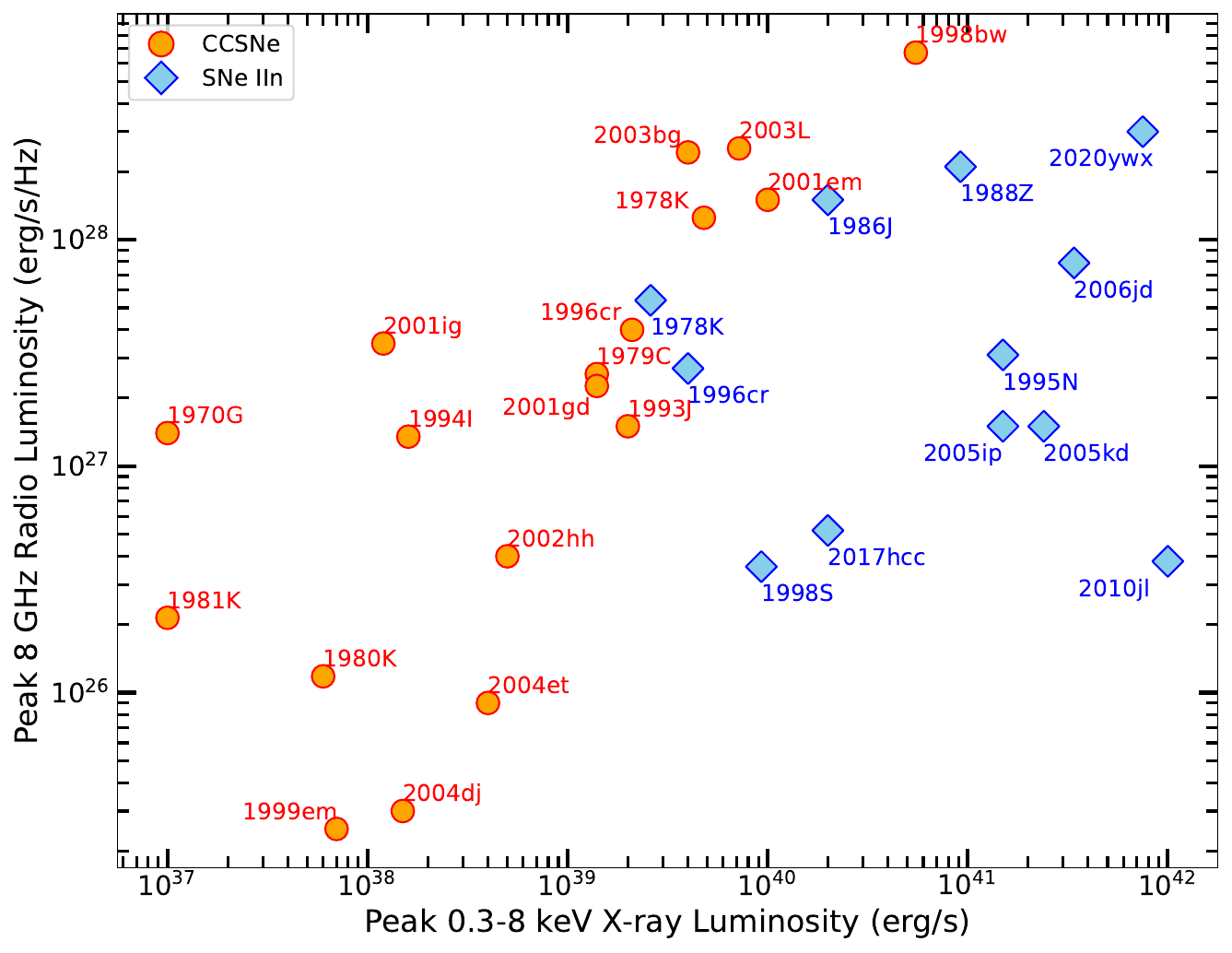}
 \caption{Radio and X-ray peak luminosities of SNe IIn compared with some well-observed CCSNe. While SNe IIn do not stand out in radio bands, they
occupy higher values in the X-ray luminosity space. The figure is modified from Chandra  \cite{Chandra2018}}
   \label{fig:iin_radioxray}
\end{center}
\end{figure}

  \begin{figure}
\begin{center}
\includegraphics[angle=0,width=0.80\textwidth]{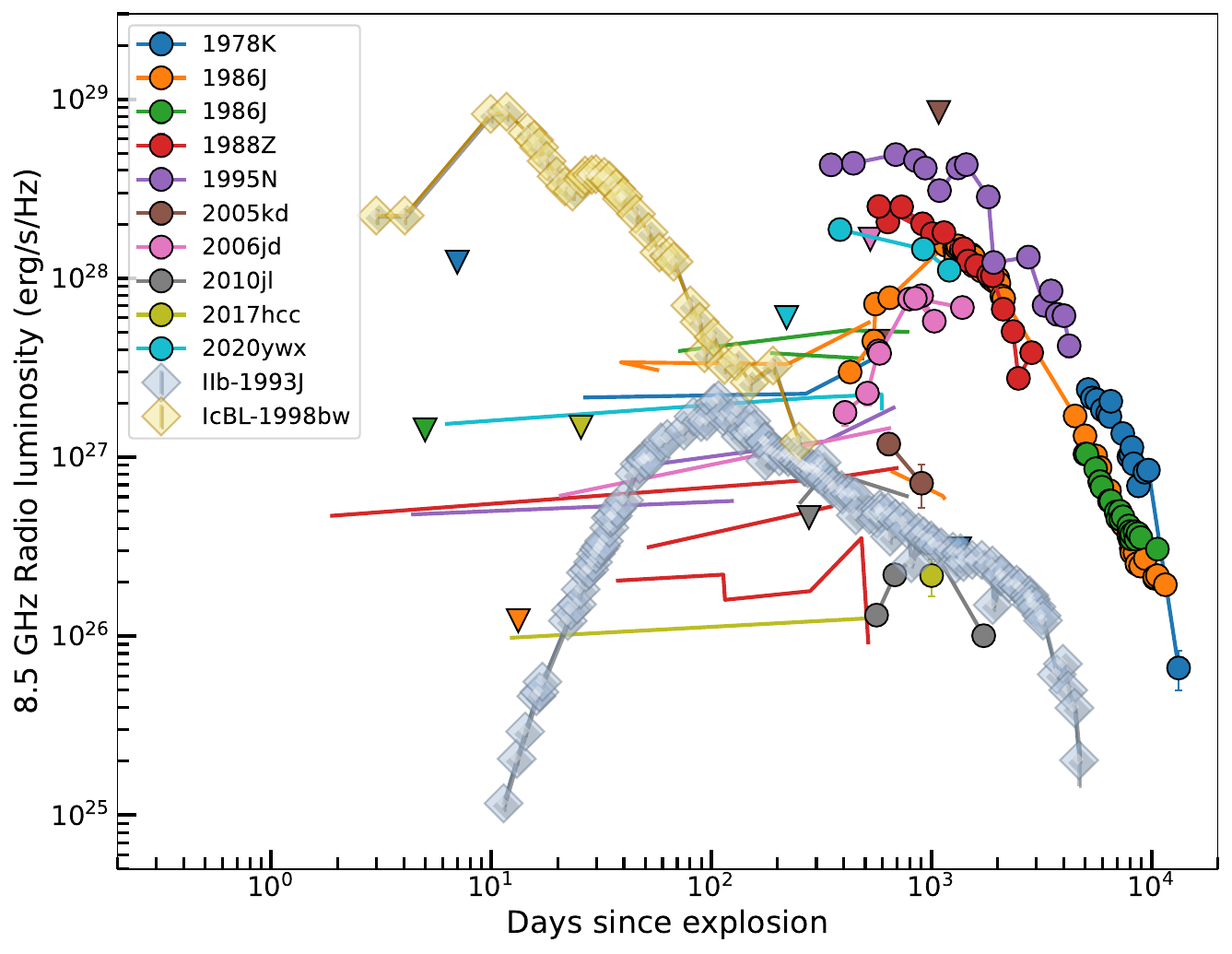}
 \caption{8 GHz radio spectral luminosity of  well-observed SNe IIn (adapted from \cite{Chandra2018}). Here, circles indicate detections. In case of non-detection, when there is only one data point, it is marked with a downward arrow. In case of multiple non-detections, only line segments have been drawn. The upper limits are 3$\sigma$ values of the map noise near the SN. We also compare the SNe IIn light curves with those of SN 1993J (IIb SN) and SN 1998bw (Ic-bl SN) to show the slower evolution of SNe IIn. }
   \label{fig:iin_lateradio}
\end{center}
\end{figure} 
 
In optical bands, evidence of  interaction manifests as  narrow ($\sim 100$\,\kms) and intermediate width ($\sim 1000$\,\kms) emission lines.  The former originates  
 from the unshocked CSM photoionized by SN radiation, and the latter
from  the CDS and/or shocked clumpy wind \cite{Chugai1994}. Additionally, broader components (thousands of \,\kms) are observed, indicative of shocked ejecta and/or broader wings due to electron scattering \cite{schlegel1990, Huang2018}.

While high CSM densities often lead to strong radio and X-ray emissions, 
only a subset of SNe IIn  have been detected   in these bands \cite{Chandra2018}. 
Some SNe IIn exhibit bright emission in both radio and X-ray emission, while others show faint radio emission despite bright X-ray emission (e.g., SN 2010jl, SN 2017hcc, also see Fig. \ref{fig:iin_radioxray}) \cite{Chandra2015, Chandra2022}. This may indicate two possible subtypes of SNe IIn.

SN 1986J \cite{Weiler1991}, SN 2006jd \cite{Chandra2012}, and SN 2020ywx \cite{Baer-way25} have shown internal FFA to be a dominant  absorption mechanism of their radio emission. Chandra et al. \cite{Chandra2012} have shown that a modest amount of cool gas ($\sim 10^4$ K) i.e., $\sim 10^{-8}\rm M_\odot$ mixing in the synchrotron emitting region can explain the observed  internal absorption.

\begin{figure}
\begin{center}
\includegraphics[angle=0,width=1.06\textwidth]{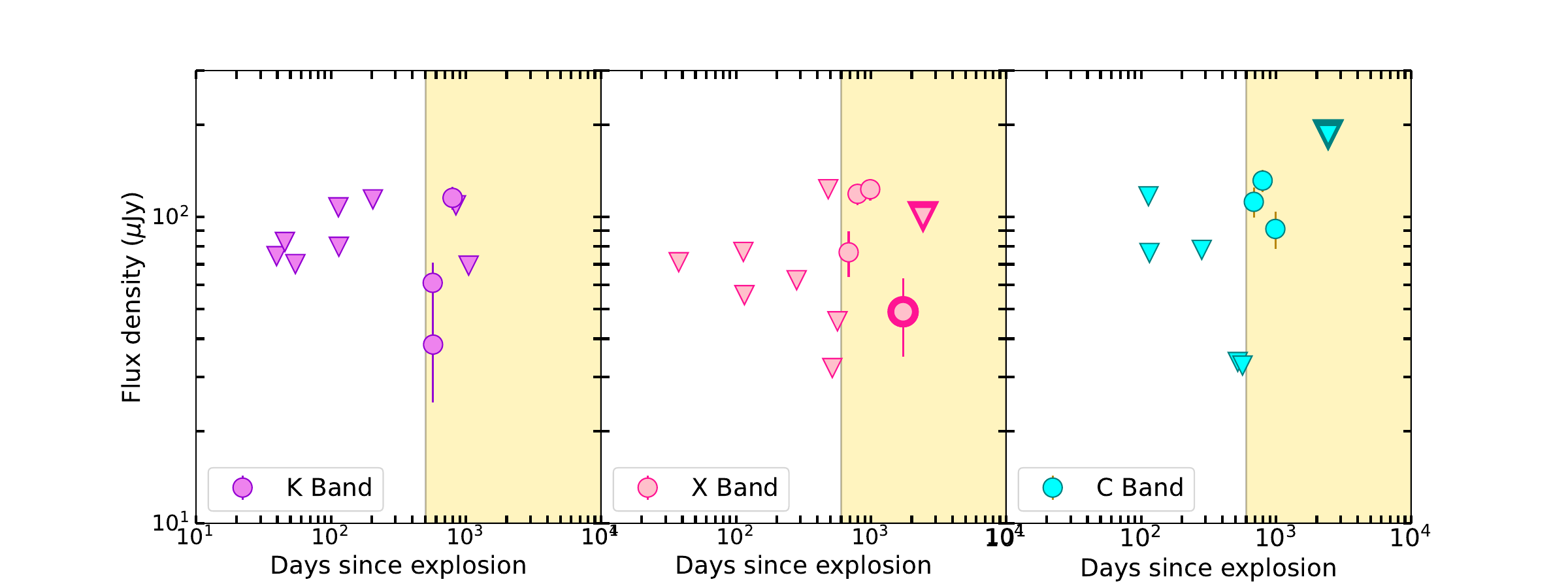}
 \caption{Radio lightcurves of SN 2010jl in VLA K (22 GHz), X (10 GHz), and C (6 GHz) bands. The yellow shaded box represents the epochs starting detections. The first radio detection started after day 500. The figure is modified from Chandra et al. \cite{Chandra2015}. The data  with bold edges are previously unpublished data points in 2015, 2016, and 2017.}
   \label{fig:sn2010jl-radio}
\end{center}
\end{figure}

SNe IIn  are   late radio emitters (Fig. \ref{fig:iin_lateradio}).
 A significant  contributing factor could be extremely high densities, leading to prolonged heavy absorption until the synchrotron emission strength has substantially decreased.
However,  some observational biases cannot be ruled out. Some SNe IIn 
were classified late and hence were not observed early enough, e.g., SN 
1995N \cite{Chandra2009_1995N}, SN 2006jd \cite{Chandra2012}, SN 1986J 
\cite{Weiler1986}. Some SNe IIn  truly are late emitters, 
such as SN 2010jl and SN 2017hcc  \cite{Chandra2015, Chandra2022}. SN 
2010jl deserves  special attention. Even though the radio observations commenced within a month, the first detection occurred on day 566 at 22 
GHz \cite{Chandra2015} (Fig. \ref{fig:sn2010jl-radio}). By this time, 
the strength of the synchrotron emission had decreased significantly so 
that SN 2010jl remained a weak radio emitter despite being one of the 
brightest X-ray SN of this class. In SN 2006gy, the radio non-detection is 
attributed to  absorption of radiation by an extremely dense medium rather than to the lack of emission \cite{Ofek2007, Chevalier2012, Svirski}. 
SN 2009ip stands out as an exception, though its radio emission faded below detection within a few tens of days \cite{Margutti+2014}. 
However, SN 2009ip was a  peculiar SN characterized by  pre-explosion outbursts and an explosion mechanism that is not yet fully understood.

Radio observations have revealed variable mass-loss rates in most SNe IIn \cite{Baer-way25, Dwarkadas+16}. In SN 2005kd, radio data indicated rapid variability in the wind mass-loss parameters in  the last 5000 years  prior to core-collapse \cite{Dwarkadas+16}. SN 2005kd was unique in other aspects as well, e.g., the optical lightcurve of SN 2005kd  showed an unusually long plateau lasting $\sim200$ days \cite{Tsvetkov2008}.
Various pieces of evidence, such as spectropolarimetry  \cite{Patat2011}, as well as radio and X-ray data \cite{Chandra2012, Chandra2018},   point towards asymmetry and/or complex geometry in SNe IIn.  Variable mass-loss rates suggest  complex mass-loss histories.  Multi-wavelength observations (X-ray, optical, infrared, and radio) in SN 2020ywx revealed inconsistencies in mass-loss rates calculated from different wavelengths, which were best explained by asymmetric CSM created due to binary interaction as  the primary mechanism for the overall mass-loss evolution \cite{Baer-way25}.

  \begin{figure}[h]
\begin{center}
\includegraphics[angle=-90,width=0.68\textwidth]{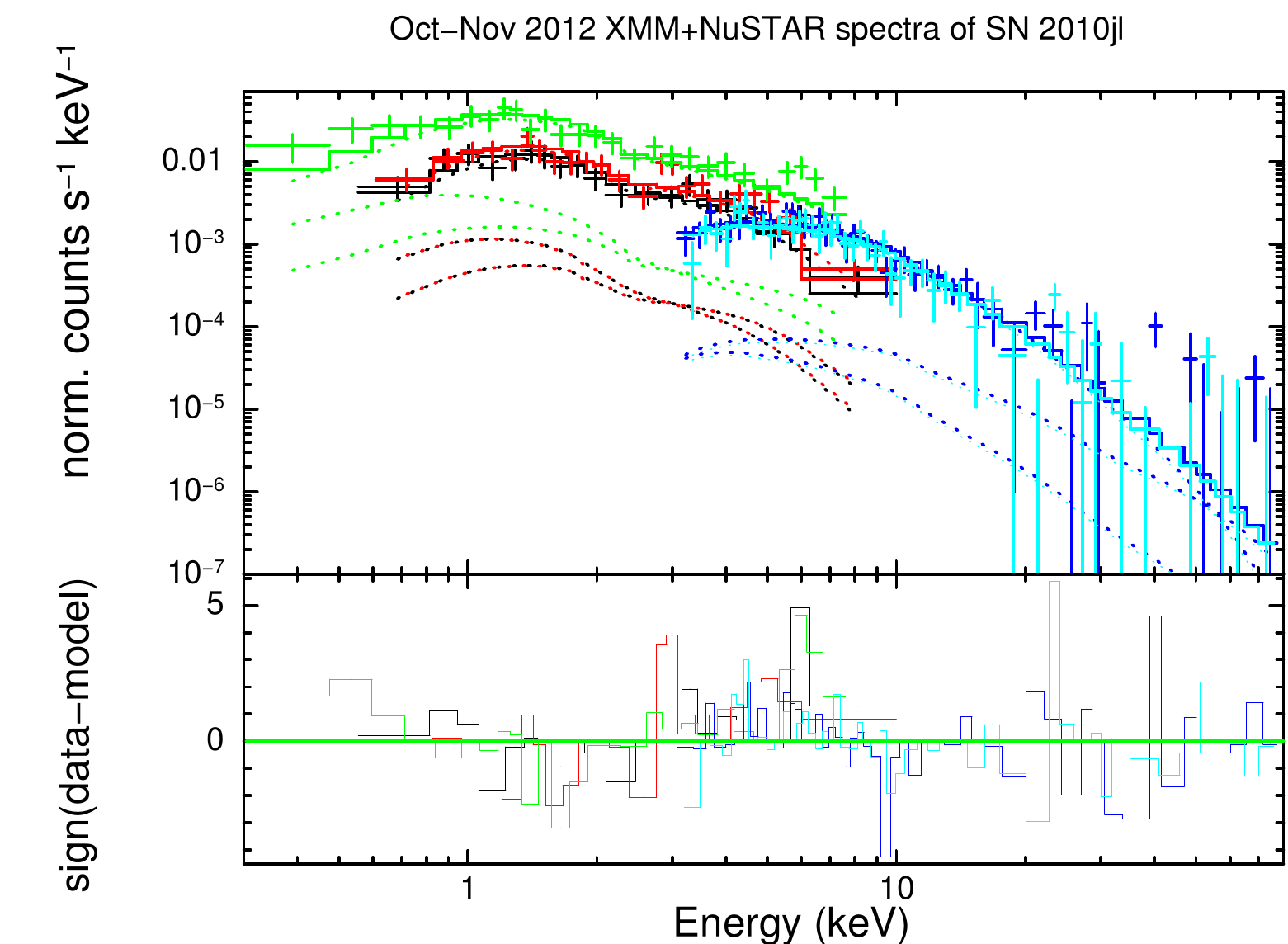}
 \caption{Joint \nustar{} and \xmm{} spectrum of the SN 2010jl covering energy range 0.3--80\,keV (reproduced from \cite{Chandra2015}).  A normalization discrepancy, attributed to the lack of sensitivity towards higher energies in \xmm{}, is evident when comparing the high-energy end of the \xmm{} spectrum with the same energy range of the \nustar{} spectrum.}  
   \label{fig:sn2010jl-xray}
\end{center}
\end{figure}

In X-ray bands, evidence of hotter FS dominance via inference of high plasma temperature has been seen in several SNe IIn, e.g., in SNe 2005ip, 2006jd, and 2020ywx \cite{katsuda+14, Chandra2012, Baer-way25}.  
However, most of these observations were made with telescopes like \chandra, \xmm, \swift, which are sensitive within 0.2--10\,keV range,  thus the plasma temperature could not be well constrained even though the 
best fit models of these SNe showed X-ray emitting plasma to be hotter than  $ 10$\,keV.  SN2010jl was the first Type IIn SN for which the exact shock temperature was determined owing to the hard X-ray sensitivity of \,\nustar. In this case, the shock temperature was determined to be 19\,keV, confirming the  FS origin of X-rays \cite{ofek+14b, Chandra2015} (Fig \ref{fig:sn2010jl-xray}).

\begin{figure}
\begin{center}
\includegraphics[angle=0,width=0.88\textwidth]{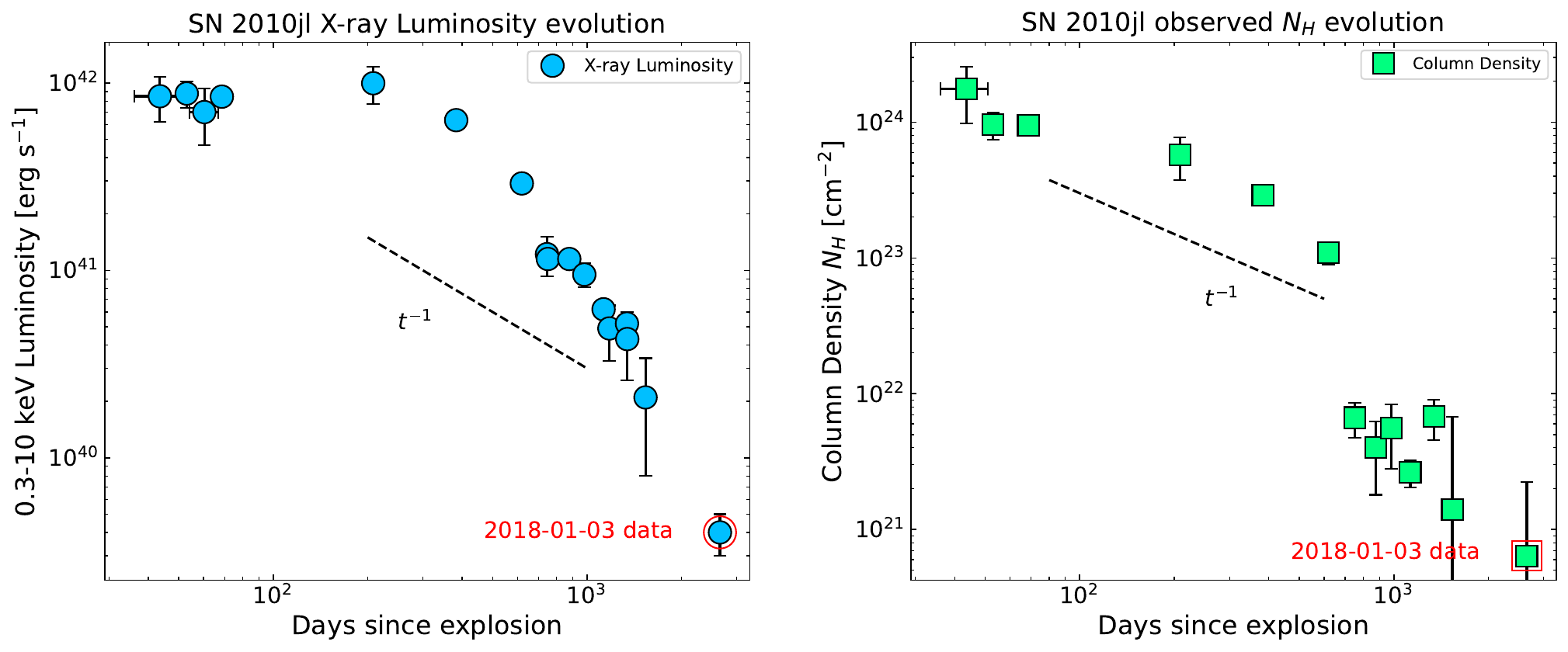}
 \caption{Left panel: Evolution of X-ray luminosity for SN 2010jl. A deviation from $t^{-1}$ is evident.   Right panel: Evolution of column density from day 40 to day 2650. The figure is modified from Chandra et al. \cite{Chandra2015}, where we have added an unpublished 46ks {\it Chandra} data on 3 Jan 2018.}
   \label{fig:sn2010jl-newdata}
\end{center}
\end{figure}

Another property of X-ray evolution in SNe IIn  is that their
X-ray light curves seem to deviate from the expected power law index of 1 (Fig. \ref{fig:sn2010jl-newdata}). This could either be due to non-smooth CSM \cite{Chugai1994, Baer-way25}, or due to  observation biases, as decreasing shock temperature at later times shifts the X-ray emission to progressively lower temperatures measured with the fixed energy range of X-ray telescopes. Most likely, the observed deviation is a consequence of both effects \cite{Chandra2012,Baer-way25}.

The direct evidence of CSM has been seen in some SNe IIn via the evolution of column density observed in X-ray observations, e.g., SN 2020ywx \cite{Baer-way25}, SN 2005ip \cite{katsuda+14}.  SN 2010jl deserves special mention due to its excellent coverage from soft to hard X-rays using  \chandra, \xmm{}, \swift{} and \nustar{}, which  allowed Chandra et al. \cite{Chandra2015} to witness three orders of magnitude
evolution in column density between day 40 and day 2600
(Fig. \ref{fig:sn2010jl-newdata}).

In Table \ref{tab:IIn_XR}, we tabulate  well-studied SNe IIn and their properties. We list  their peak X-ray and 8 GHz spectral radio luminosities and the time to reach the peak in both wavebands. We also compile the column density and plasma temperature from the literature, along with the epoch of first radio detection.

\begin{table}[]
  \small    
    \centering
  \begin{threeparttable}
    \caption{X-ray and Radio Properties of Type IIn Supernovae}
    \label{tab:IIn_XR}
    \begin{tabular}{llllllllll}
      \toprule
            & \multicolumn{4}{c}{\textbf{X-ray}} &
              \multicolumn{4}{c}{\textbf{Radio}} &
              \\ 
    \cmidrule(lr){2-5} \cmidrule(lr){6-9}
    SN  & $L_{\mathrm{p}}^{X}$ & $\Delta t_{\rm peak,X}$
    & $N_H$ & $kT$ & $\Delta t_{\rm det,R}$ &
    $\nu_{\rm det}$ & $L_{\nu,\mathrm{p}}^R$ & $\Delta t_{\rm p,R}$ & Ref.\\
      & erg\,s$^{-1}$ & days  &cm$^{-2}$ & keV & days  & GHz & erg\,s$^{-1}$\,Hz$^{-1}$ & days & Ref.\\
    \midrule
     1978K  & $ 2.6 \times10^{39}$ & 5600 & $2.1\times10^{21}$ & 3.3 & $1300$ & 0.84 &
      $5.4\times10^{27}$ & $\sim$2740 &
  \cite{Ryder1993_78K,Schlegel1999_78K,Chiba2020_78K,Smith+2007} \\
        1986J & 
         $2 \times10^{40}$ & 4150 &  $\sim 10^{22}$ & 6  & 1490  & 4.86 & $1.5\times10^{28}$ & 2225 & \cite{weiler1990}\\
        1988Z  & $ 9.2 \times10^{40}$ & 2376 & $5\times10^{21}$
        & $2-3$ & 385 & 4.86 & $2.1\times10^{28}$ & 743 & \cite{Williams2002, Schlegel2006}\\
        1995N  & $1.5\times10^{41}$ & 1477 & $1.1\times 10^{21}$ & $9.1$ & 350 & 8.46 & $3.1\times10^{27}$ & 688 &\cite{Chandra2009_1995N, Chandra2005}\\
        1996cr  & $4\times10^{39}$&  3770 & $7.7\times10^{21}$ & 25 & 397 & 8.46 & $2.7\times 10^{27}$ & 3198 
        & \cite{Bauer2008_SN1996cr, Patnaude2025, Meunier2013}\\
        1998S & $9.3\times10^{39}$ & 678 & $1.4\times10^{21}$ & 9.8 &
        310 & 8.46 & $3.6\times 10^{26}$ & $\sim1000$& \cite{Pooley2002_SN1999em}\\
        2005ip   & $1.5\times10^{41}$& 784 &  
        $1\sim10^{22}$ &$20$ &$<3743$ & 8.46 & $1.5\times10^{27}$ &$<3743$ 
         & \cite{katsuda+14,smith2017}\\
        2005kd & $2.4\times10^{41}$ & 479 & $\sim 10^{22}$ &$\ge10$ & $<632$ & 8.5& $1.5\times10^{27}$ &$<632$ 
         & \cite{Dwarkadas+16}\\
         2006jd &  $3.4\times10^{41}$ & 403 & $1.8\times 10^{21}$ & $\ge20$ & 405 & 8.46 &
         $7.9\times10^{27}$ & 902 & \cite{Chandra2012}\\
         2010jl  & $1\times10^{42}$ & 208 & $10^{24}$ $\rightarrow$
         $10^{21}$ & 18 & 566 & 21 & $3.8\times10^{26}$ & 793 & \cite{Chandra2015}\\
        2017hcc  & $2\times10^{40}$ & 727 & $\sim 10^{22}$ & $3-20$ & 1002 & 10 & $5.2\times10^{26}$ &
        1003 & \cite{Chandra2022}\\
        2020ywx  & $7.5 \times10^{41}$ & 230 & $4\times10^{22}$ &$\sim20$ & 256 & 1.4 & $3\times10^{28}$ & 990 & \cite{Baer-way25}\\
      \bottomrule
    \end{tabular}
    \begin{tablenotes}
      \footnotesize
      \item \textbf{Notes.}
      \item $\Delta t_{\rm dec,R}$ is  days since explosion to first radio detection.
      \item Peak radio spectral luminosity is at 8 GHz.
    \item Luminosity measurements have at least 10\% errors.
    \end{tablenotes}
  \end{threeparttable}
\end{table}

  \begin{figure}
\begin{center}
\includegraphics[angle=0,width=0.99\textwidth]{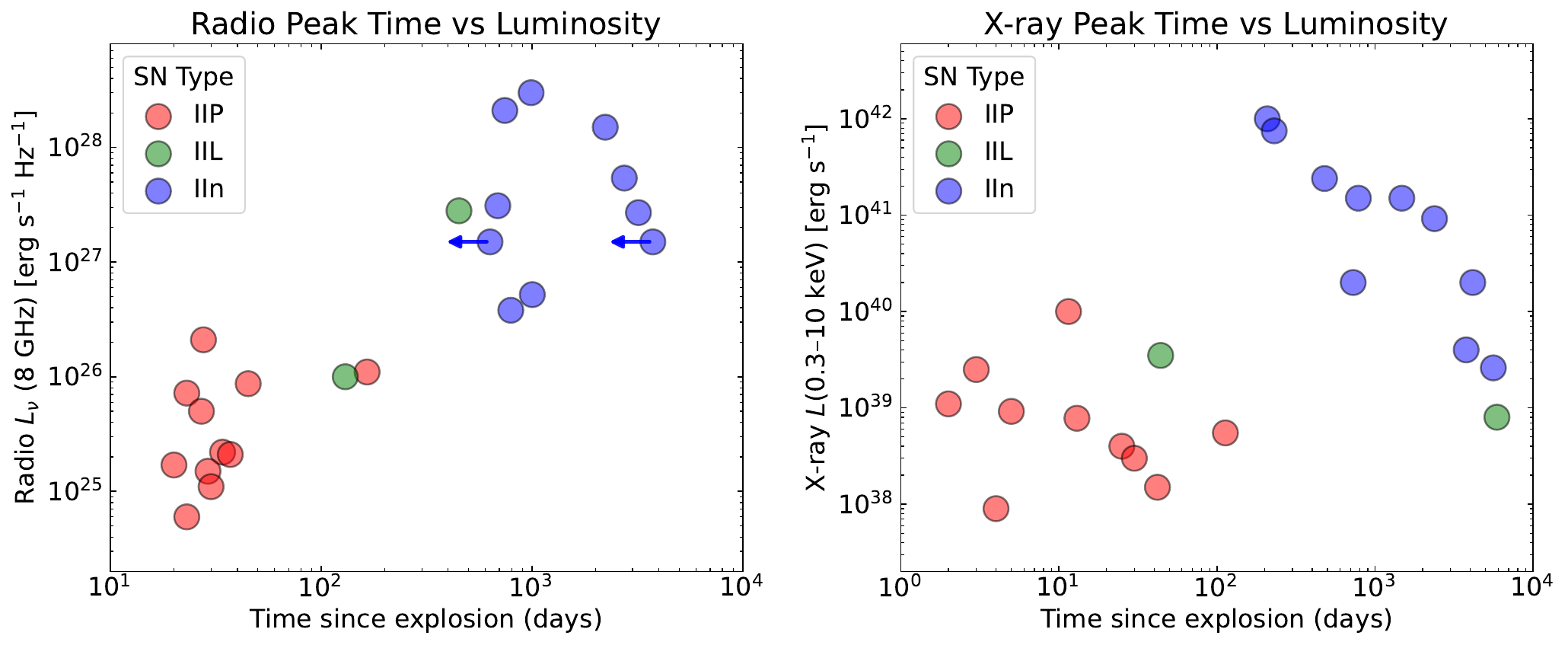}
 \caption{Left: Comparison of peak Radio luminosity and time to peak for SNe IIn vs SNe IIP/IIL/II. Right: Comparison of peak X-ray luminosity and time to peak for SNeIIn vs IIP/IIL/II.  The plots indicate  a distinct population of these two classes of SNe, despite both classes being H-rich. See Table \ref{tab:IIn_XR} for references.}
   \label{fig:IIP-IIn}
\end{center}
\end{figure}

Even though SNe IIn are H-rich SNe, they seem to possess significantly different properties than their non-interacting counterparts. 
In Fig. \ref{fig:IIP-IIn}, we plot peak radio and X-ray luminosities of H-rich (IIP, IIL, and II SNe) along with IIn SNe.  While both are H-rich SNe, 
SNe IIn are systematically brighter and peak later, whereas SNe IIP peak earlier and are fainter. This most likely indicates two very distinct populations giving rise to SNe II/IIP/IIL and SNe IIn.

\paragraph{SN 2010jl at $\sim 7$ years post explosion}

SN 2010jl is one of the best studied SNe IIn.
Ofek et al. \cite{ofek+14b} presented early optical and X-ray data, whereas Chandra et al. \cite{Chandra2015} carried out comprehensive radio and X-ray analysis up to day 1500. Here we extend the observation baseline to $\sim 2500$\,d by adding the unpublished  X-ray data taken 
on January 3, 2018, and 
radio data (with the VLA) from June 23, 2015, to May 23, 2017. This X-ray observation was conducted using  \chandra{} (Proposal \#18500302, PI: Chandra) with a 46 ks exposure with ACIS-S, yielding 35 counts.
 We analyzed the spectrum using the C-statistic and fit an absorbed thermal plasma model with metallicity of 0.3, as obtained from Chandra et al. \cite{Chandra2015}. Due to the low number of counts, fitting for all three parameters (column density, temperature, and normalization) did not produce a meaningful fit. Therefore, we initially fixed the column density to $10^{21}$\,cm$^{-2}$ (last measured value from \cite{Chandra2015}) and varied the plasma temperature to determine if the X-ray emission was dominated by the RS at late epochs. Our best fit temperature,  $16\pm70$\,keV,  is not well-constrained but suggests a continuing possible FS origin. Consequently, we froze the temperature at 19 keV (as reported by Chandra et al. \cite{Chandra2015}) and fit  the column density. 
 The best fit column density is 
$N_H=(0.6\pm1.6)\times10^{21}$\,\cm. The absorbed (unabsorbed) flux in 0.3--10 keV is $(1.31\pm0.36)\times10^{-14}$\,\ergcms{} ($(1.41\pm0.39)\times10^{-14}$\,\ergcms). This translates to 0.3--10 keV unabsorbed luminosity of $\sim 4\times10^{39}$\,\ergs{} on day 2651. Chandra et al \cite{Chandra2015}  reported a  luminosity of  $\sim 2\times10^{40}$\,erg\,s${-1}$, on day 1530, indicating a luminosity decline proportional to $t^{-2}$ between the two epochs.
 The column density has now reached that of the Galactic column density (Fig. \ref{fig:sn2010jl-newdata}. These findings are consistent with SN 2010jl having emerged from its  CSM by this epoch. Given shock and wind speeds of 4000\,km\,s$^{-1}$ and 100\,km\,s$^{-1}$, respectively \cite{Chandra2015}, this implies a pre-explosion interaction period of approximately 300 years. This suggests that the enhanced mass-loss rate of 0.1\,M$_\odot$\,yr$^{-1}$, as reported by Chandra et al. \cite{Chandra2015},  commenced after this epoch in the SN 2010jl progenitor.

The radio observations were conducted in VLA S (2--4\,GHz), C (4--8\,GHz), X (8--12\,GHz) and Ku (12--18\,GHz) bands in 2015, Ku band in 2016 and S, C and X bands in 2017 (Table \ref{tab:radio_10jl}). The 2015 
and 2016 observations were in the highest A configuration of the VLA and resulted in detections at all frequencies, albeit with lower flux 
densities than the value reported by Chandra et al. \cite{Chandra2015} (all see Fig. \ref{fig:sn2010jl-newdata}), ruling out any dense shell. The 2017 observations were at a lower-resolution C configuration and were contaminated by nearby radio bright sources, resulting in upper limits.

\begin{table}
\centering
\caption{SN 2010jl late time VLA radio observations}
\label{tab:radio_10jl}
\begin{tabular}{lcccc}
\hline
Date of obs. &  Days since & Freq & VLA  & Flux density\\
 & expl. & GHz & config. &$\mu$Jy   \\
\hline
23 Jun 2015	&1726	&10	&A	&48.9$\pm$14.1\\
28 Jun 2015	&1731	&6	&A	&37.2$\pm$9.9\\
17 Jul 2015	&1750	&15	&A	&35.0$\pm$11.5\\
15 Aug 2015	&1779	&3	&A	&59.5$\pm$23.2\\
17 Nov 2016	&2239	&15	&A	&32.5$\pm$11.7\\
21 May 2017	&2424	&3	&C	&$<298.7$\\
21 May 2017	&2424	&10	&C	&$<99.8$\\
23 May 2017	&2426	&6	&C	&$<185.6$	\\
\hline
\end{tabular}
\footnotesize
\caption*{\textit{Notes. Upper limits are 3-$\sigma$.
Date of explosion is assumed to be 1 Oct 2010 \cite{Chandra2015}. The low resolution data in VLA C configuration is contaminated by nearby radio sources resulting in higher upper limit.}}
\end{table}

 \subsubsection{Type Ibn/Icn supernovae}

SNe Ibn/Icn generally show a rapid decay in their light curves  after the peak, unlike slowly evolving SNe IIn.  This suggests that in SNe Ibn/Icn, the denser CSM extends to smaller radii. 
Due to high-density CSM, the spectra taken within 1--3 days  usually show the flash-ionization signatures, e.g., SN 2019uo \cite{Gangopadhyay2020}, SN 2023emg \cite{Pursiainen2023} among SNe Ibn and SN 2023xgo \cite{Gangopadhyay2025} among SNe Icn. Dust formation signatures due to high density have been seen in some SNe, e.g., SN 2006jc \cite{Mattila2008,Smith2008}.

The signatures of CS interaction in radio and X-ray bands are rare in SNe Ibn/Icn due to the rarity of these events and rapid evolution. 
 The only reported X-ray detection from SNe Ibn are SN 2006jc \cite{Immler2008} and SN 2022ablq \cite{Pellegrino2024}.
 The \swift{} observations of SN 2006jc during 19--183 days revealed X-ray emission with luminosity $10^{39}$\,\ergs{}
 \cite{Immler2008}.  
SN 2022ablq was found to be more than an order of magnitude brighter than SN 2006jc \cite{ Pellegrino2024}. In addition, SN 2010al, which was initially classified as IIn but later proposed to be a Ibn SN, has shown detectable
X-ray emission \cite{Ofek2013}.
 No SNe Icn has been detected in X-ray bands so far. SN 2019hgp was  observed with the \swift{} during the first two months, but no detection was reported \cite{Galyam2022}.
 Inoue et al. \cite{Inoue2024} indicated that important progenitor properties can be derived based on X-ray modeling of these SNe.
By analysing X-ray data of the above three SNe, they concluded SN 2022ablq was more stripped than SN 2006jc and that the prior SN contained a smaller fraction of C and O.

SN 2023fyq is the first Ibn SN from which radio emission has been seen, and it is consistent with a merger scenario leading to a short-term mass-loss before the explosion \cite{Baer-Way2025b}.
 
\subsubsection{Supernovae undergoing metamorphosis}

Some SNe, which are initially classified differently, metamorphose into interacting SNe. Observations of these SNe usually reveal key aspects about their progenitor stars.
The most notable SN undergoing  metamorphosis
is SN 2014C. Initially, it was classified as a  Type Ib SN, which later showed interaction signatures via strong H-$\alpha$ emission, leading to IIn classification \cite{Milisavljevic2015}. 
AMI-LA high cadence observations at 15 GHz 
revealed a rise in two stages, with the first peak of 0.5 mJy occurring at around 80 days, and then a second peak of 2 mJy at around 250--300 days \cite{Anderson2017_SN2014C_AMI}.
AMI-LA light curve, therefore, provided the clearest  evidence that SN 2014C exploded in a low-density bubble and then “metamorphosed” into a IIn SN when it struck a massive, H-rich shell ejected decades earlier. 
Margutti et al.  \cite{Margutti2017} presented \swift, \chandra{}, and \nustar{} data up to around 1000 days. They found the shock temperature to be evolving from 18--20 keV to 7--8\,keV and column density evolving from $3\times10^{22}$\,cm$^{-2}$ to  $5\times10^{21}$\,cm$^{-2}$. Later, Thomas et al. \cite{Thomas+22} and Brethauer et al. \cite{Brethauer+22} presented 7 years of \chandra{} and
\nustar{} datasets. Their work confirmed the FS origin of the X-ray emission, later revealing  smaller temperatures, possibly from the RS. Both works remain the only ones to present a 7-year hard X-ray data set for a SN, in addition to the Type IIn SN 2010jl presented here.
The radio and X-ray observations were best explained in a scenario in which the SN 2014C progenitor had lost its hydrogen envelope close to the explosion, and then in about 100 days the shock encountered the lost hydrogen envelope of around 1\,\msun{} at $5\times10^{16}$\,cm away. 

SN 2001em was another  SN undergoing metamorphosis. Initially, it was classified as Type Ib/c SN and later reclassified as type IIn when Stockdale et al. \cite{Stockdale_2001em}
detected radio emission with the VLA two years after the explosion. Pooley et al. \cite{Pooley2002_SN1999em} also detected X-ray emission with \chandra{} with X-ray luminosity reaching $\sim 10^{41}$\,\ergs. While initially the radio and X-ray detection was identified as possibly coming from an  off-axis GRB associated with the SN \cite{Granot2004}, the VLBI observations found the expansion speed to be non-relativistic \cite{VLBA_2001em}, and it was reclassified as SN IIn with delayed CSM interaction. 
Chugai \& Chevalier \cite{Chugai2006_SN2001em} modeled the emission coming from a  dense CSM ring   and indicated that  the radio and X-ray emission could be attributed to  interaction of the SN  ejecta with a 3\,M$_\odot$  CS shell at a distance of $7\times10^{16}$\,cm. They interpreted that the hydrogen CS shell was formed as a result of enhanced mass-loss rate ($\dot M=(2-10)\times10^{-3}$\,\ml) around 1000 yrs prior to the SN explosion. Their observations covered 1000 days, and the modeling indicated that the ejecta hadn't reached the edge of the shell by the end of the observed duration. Later, Chandra et al. \cite{Chandra2020} presented  a 19-year timeline of SN 2001em. From X-ray temporal behavior, they found that the shock exited the shell at the latest by  1750 days,  and entered a faster, lower-density wind. However, late time ($\sim6000$ days) spectral inversion was seen in radio data analogous to SN 1986J \cite{bietenholz+17}.  Chandra 
et al. \cite{Chandra2020} interpreted that the  spectral inversion was due to  a slower shock  inside the faster shock, possibly formed in the equatorial regions created due to  binary interaction. These observations were best explained by a massive binary system with a common-envelope phase \cite{Chandra2020}.
If this interpretation is correct, SN 2001em is the only SN in which an outer CS shell (duration 1000--1750 days) and an inner slower shock ( at $\ge6000$ days) have been revealed. This gives a unique perspective on the binary evolution of the SN progenitor star.

Other examples in this SNe undergoing metamorphosis into interacting class are  SN 2018ijp, SN 2017dio, SN 2004dk, SN 2019oys (see Table \ref{tab:meta} for list and references).
SN 2018ijp and SN 2017dio were initially classified as Ic SN, but showed emergence of narrow H lines. The strong CSM interaction was revealed months after the explosion in SN 2018ijp \cite{Tartaglia2021}, but almost immediately in the latter \cite{Kuncarayakti2018}. SNe 2004dk and 2019oys were initially classified as SNe Ib, which metamorphosed into SNe IIn. While this happened within months in SN 2019oys \cite{Sollerman2020}, this metamorphosis was seen 13 years later in SN 2004dk \cite{Mauerhan2018}. No detectable  radio or X-ray 
emissions were seen from SN 2017dio and SN 2018ijp; however, SN 2004dk and SN 2019oys were seen to be bright in both radio and X-ray emissions. These cases reveal delayed CSM interaction, pointing to non-steady and complex mass-loss histories of massive stars shortly before explosion. 
The table \ref{tab:meta} provides a summary of these SNe.

\begin{table}[ht]
\centering
\caption{List of Supernovae undergoing metamorphosis to SNe IIn}
\begin{tabular}{llllccl}
\hline
& & \multicolumn{2}{c}{\textbf{Interaction signatures}} &
\multicolumn{2}{c}{\textbf{Interaction signatures}} & \\
 \cmidrule(lr){3-4}
\textbf{SN} & \textbf{Initial Type} & \textbf{$\Delta t$} & Bands & \textbf{Radio} & \textbf{X-ray}& \textbf{Ref.} \\
 & Type & days & & & & \\
\hline
 2001em & Ib/c & $\sim2$\,yrs & radio & yes & yes & \cite{Chugai2006_SN2001em, Chandra2020}\\
2004dk & Ic & 13 yrs & optical & yes & yes & \cite{Aldering2006_2005gj,Mauerhan2018,Pooley2019} \\
2014C & Ib & $\sim 100$d & R, O, X & yes & yes  & \cite{Milisavljevic2015} \\
2017dio & Ic & almost immediately & optical & no & no & \cite{Kuncarayakti2018} \\
2018ijp & Ic & months & optical & no & no &  \cite{Tartaglia2021} \\
2019oys & Ib & months & optical & yes & yes &  \cite{Sollerman2020} \\
\hline
\end{tabular}
\label{tab:meta}
\end{table}

There are also SNe that were never reclassified as SNe IIn, but they have shown late-time rebrightening due to interaction. 
A SN which deserves a special mention in this category is SN 2018ivc. It was classified as a IIb/IIL \cite{Bostroem2020}, but around $\sim1000$ days, it revealed dramatic rebrightening with the VLA, e-MERLIN, and the ALMA \cite{Mutie2022, Maeda2023}. Maeda et al. \cite{Maeda2023}
interpreted a high mass-loss phase ($10^{-4}$\,\ml)  1500\,yr prior to explosion, followed by an order of magnitude lower mass-loss phase until the explosion.
Some other examples in this category are SNe  2004gq \cite{Nagy2025}, SN 2001ig \cite{Soria2025}
and the oldest example SN 1970G \cite{Dittmann2014}.
In data consisting of 2 decades post-explosion, SN 2001ig has shown two orders of magnitude larger rebrightening  than expected in the standard CS interaction model \cite{Soria2025}.
This suggests that
the SN ejecta have reached a denser shell, perhaps compressed by the fast wind of a WR progenitor or expelled centuries before
the final stellar collapse or due to binary interaction.

 \subsection{Superluminous supernovae}

SLSNe classification is purely an observational one. The majority of them are expected to reach their high luminosities via mechanisms involving the central engine, CS interaction, or radioactive decay. 
SLSNe have mostly eluded radio and X-ray emission.
Surveys to search for radio emission have mostly obtained null results
 \cite{Hatsukade2021}.
So far, only two SLSNe (both SLSNe I) have shown radio emission,  PTF10hgi 
\cite{Eftekhari2019, Eftekhari2021, Mondal2020} and  SN 2017ens \cite{Margutti2023}. 
There was a claim for the third SLSN  SN 2020tcw  to be radio bright; however,  follow-up observations revealed the radio emission to be arising from  an unrelated source
\cite{Margutti2023}.  
In PTF10hgi, radio emission was discovered 7 years after the explosion, and radio modeling revealed that the SN was  powered by a magnetar and radio emission was arising from the pulsar wind nebula surrounded by CSM.  SN 2017ens initially showed Balmer series H-lines emission in their spectra after 100 days \cite{Chen2018} and radio detection (at 3 years post explosion) revealed CS interaction to be the main power source.  
 Margutti et al. \cite{Margutti2023} inferred a mass loss rate of 
$\dot M \approx 10^{-4}-10^{-3}$\,\ml{} for a wind of 1000\,\kms.

Amongst CS interaction-powered SLSNe, SN 2006gy deserves special mention. While it eluded radio emission due to absorption of radio emission via  high density CSM, and resulted in weak X-ray emission \cite{Smith+2007} due to X-ray suppression via multiple effects caused by optically thick CSM \cite{Chevalier2012}, the bright optical emission was result of the shock propagating into optically thick CSM and converting kinetic energy of the shock into radiation \cite{Chevalier2011}.


While no radio or X-ray emission has been claimed from SLSNe I with the PPISNe  pathway, Lunnan et al. \cite{Lunnan2018} found the first clear UV resonance-line light echo (in Mg II) from a CS shell around the type I SLSN iPTF16eh. They modeled the time-variable absorption and emission with a shell at $\sim 0.1$ pc expanding at 3300\,\kms, which was ejected 30 years before the explosion. They claimed that these properties were consistent with a PPISNe ejection from a progenitor with a $\sim 50$\,\msun{} He-core (ZAMS mass $\sim 115$\,\msun). 
In another PPISNe candidate, iPTF14hls, late time ($>1000$ d) spectrum revealed 
a double-peaked intermediate-width H$\alpha$ line corresponding to expansion speed around 100\,\kms, which was a clear signature of CS interaction with a possible disc-like CSM \cite{Andrews2018}.

 \subsection{Electron capture supernovae and Calcium-rich supernovae}

 The primary example of an ECSNe candidate is SN 1054, also known as the Crab \cite{Nomoto1982a}, which is an SNR. While no SN has been unequivocally identified as ECSNe, SN 2008S was argued to have an electron-capture origin,  and also an object falling under  Intermediate-Luminosity Red Transient (ILRT) - purely  an observational class with cool, dusty, intermediate-luminosity eruptions/explosions \cite{Botticella2009}. More recently, SN 2018zd in the galaxy NGC 2146 has been argued to be the best observed candidate for a ECSNe \cite{Hiramatsu2021_SN2018zd}.
Flash signatures were seen in SN 2008zd extending beyond 9 days, indicating CS Interaction \cite{Hiramatsu2021_SN2018zd}.

Among Ca-rich SNe, so far only two SNe have been reported to show X-ray emission. These are  SN 2019ehk \cite{Jacobson2020} and SN 2021gno \cite{Jacobson2022}. The X-ray emission from SN 2019ehk was seen only in the first 4 days, with peak 0.3--10 keV luminosity $L_x\approx  10^{41}$\,\ergs. The X-ray luminosity declined fast $L_x(t)\propto t^{-3}$, and was predicted to arise from  CS interaction, which was also confirmed with flash ionization features seen in the first 1.5 days \cite{Jacobson2020}. The SN was not detected in X-ray bands. 
SN 2021gno reached peak 0.3--10 keV luminosity $L_x\approx  5\times 10^{41}$\,\ergs, within a day \cite{Jacobson2022}. Combined with radio non-detections, Jacobson-Gal\'an \cite{Jacobson2022}
inferred the extent of CSM to be $<3\times10^{14}$\,cm. Both SNe were consistent with mass-loss rate of $\dot M\approx 10^{-3}-10^{-2}$\,\ml. In both SNe, the derived progenitor CSM density was found to be consistent with the merger of low-mass, hybrid WDs \cite{Jacobson2022}.
 

\section{Discussion and Conclusions}

This work is focused on reviewing  the CS interaction in various kinds of SNe. We  provide an updated  diagram for 
SNe classification, incorporating stellar death pathways along with the traditional classification (Fig. \ref{fig:SNclassification}).

We argue that CS interaction is the best indicator of the diversity of progenitor channels of various types of SNe. X-ray observations provide insights into shock temperature and density, and measure important parameters like column density, mass-loss rate, etc. Radio  measurements provide information on shock velocity, size, magnetic field, and CSM density. We note that both methods provide a ratio of mass-loss rate to wind speed; thus, obtaining wind speed is important for accurate mass-loss rate values. Optical observations are the best indicators of wind speeds. However, it is possible that the SN is asymmetric and wind speeds are different in different directions (like SESNe). In addition, in some cases, wind speeds may vary, similar to WR progenitors with fast winds followed by slow winds.  Non-spherical CSM may lead to different wind speeds at different lines of sight.  Some information about the CSM winds and their complexities can also be obtained from massive and evolved stars in the
Milky Way. For example,  
RSGs and YSGs  with larger radii usually show slower winds
($ \sim10-20$\,\kms),
whereas LBVs and BSGs with
smaller radii have faster winds ($\sim100$\.\kms). In reality, though, high
radiation from the SN can potentially accelerate the CSM winds to much
higher speeds  than their pre-explosion progenitors \cite{Smith2014}.
It is important to  keep caveats
in mind while interpreting the data and getting information on the progenitors of different kinds of SNe.

We discuss various classes of SNe in the context of CS interaction. Among the thermonuclear class, we focus on SNe Ia (of any subtype) and SNe Ia-CSM. The rarity of SNe Ia-CSM makes it difficult to understand the nature of their CS interaction.
Amongst CCSNe, we divide the discussion into  hydrogen-rich SNe  (SNe IIP, IIL, II), SESNe (IIb, Ib, Ic), and  interacting SNe (IIn, Ibn, Icn, "metamorphosis" interacting SNe). We also briefly discuss observed signatures of CS Interaction in SLSNe,  ECSNe, and Ca-rich SNe. 

 Below, we summarize the main takeaways of this review.

\begin{enumerate}
    \item 
    The earliest signatures of CS interaction appear as flash ionization signatures, which typically start within hours to days and last for a few days. 
Several  H-rich SNe show  flash ionization features within $10^{15}$\,cm with mass-loss rates ranging
$\sim 10^{-4}-10^{-2}$\,\ml, indicating presence of confined CSM. This implies enhanced mass-loss activity in RSG progenitors of these SNe.  
\item While flash spectroscopy has been majorly seen in  H-rich SNe, SESNe generally have not been observed to reveal it, except SN 2013cu, which was a IIb SN, a bridge between H-poor and H-rich SN \cite{Galyam2014}.
A reason for the lack of flash spectra in SESNe is a smaller ratio of ejecta to wind speeds. This can be explained via the example of Ic-bl SN 2020oi with an ejecta to wind speed ratio of $\sim10$. 
Confined CSM   revealed with ALMA within  $10^{15}$ translated into 
the enhanced mass-loss activity in the last year, which is easy to miss in observations.   More studies are needed in this direction.
\item  Radio studies of H-rich SNe  usually show FFA or a combination of FFA and SSA to be the dominant absorption mechanism. The X-ray emission is dominated by the RS.   Due to the additional supply of photons during the photospheric phase, signatures of cooling and IC emission have been seen in radio and X-ray bands, respectively.
\item SESNe show comparatively faster evolution than their H-rich counterparts, pointing to  relatively compact progenitors.
SESNe CS interaction signatures challenge the view that they explode into clean environments, and suggest a clumpy or asymmetric CSM, possibly associated with binary evolution or eruptive mass loss. SSA is the dominant absorption mechanism of the observed radio emission. X-ray spectra usually show contributions from non-thermal emission.
\item Radio observations are not very discriminating between cIIb and eIIb subclasses in SNe IIb. This likely indicates a lack of a clear boundary between the two kinds of progenitors, suggesting a continuum in their progenitor sizes.
\item Among SESNe,  GRB-SNe evolve faster than the rest of the population.
SNe Ic-bl show diversity, and
the CS interaction is poorly understood; the contribution of the central engine is not well constrained,  though efforts are underway.
Radio observations combined with X-ray measurements are likely to be instrumental in understanding them.
\item Late time rebrightening has been seen in several SESNe or SESNe with 'metamorphosis'.  Binarity is the most plausible explanation. 
A detailed and high cadence wide band study  of SESNe in radio bands is crucial to  pinpoint the  nature of their progenitor systems.
\item It  would be worthwhile to mention that some of the non-IIn interacting subtypes (delayed interaction/some SESNe) lack optical signatures and/or show discrepancies with radio/X-ray results. This could be due to different geometries. However, it is worth mentioning that  optical observations are a unique probe of CSM speeds, which is a major diagnostic for estimating accurate mass-loss rates, which makes interpreting the SESNe with no optical signatures more difficult.
\item Amongst interacting SNe, SNe IIn show strong and persistent 
interaction with their dense CSM. Their likely progenitors, based on mass-loss rates,  are compatible with LBVs or RSGs with extreme mass-loss rates. However, it is very difficult to understand the true nature of their  progenitors, which  are usually hidden inside the optically thick dense winds. 
\item SNe IIn are late radio emitters and the slowest evolving SNe. While some of it may be attributed to observational biases, this points towards the fact that the dense CSM is absorbing the majority of the emission and is a major factor determining  the SN dynamics and evolution in this class.
\item H-poor interacting SNe, SNe Ibn and Icn,  are also surrounded by dense CSM; however, only a handful have been detected. More efforts are needed to increase their sample size and understand this class. 
\item Comparing the peak radio luminosities and the X-ray light curves of SNe II, we find that    SNe IIn usually  have the most luminous and long-lived X-ray and radio signatures. They also represent a distinct progenitor class from the rest of the H-rich SNe.
Amongst H-rich SNe, type IIL evolve faster than type IIP, consistent with a smaller H-envelope seen in optical bands.
  \item SNe that metamorphose into interacting SNe reveal similar properties as SNe IIn. Both show short cooling time scales, formation of CDS, dominance of hard X-rays from the FS, and evolution of column density.
  \item  PISNe, PPISNe, SLSNe, ECSNe, Ca-rich SNe, etc., are very poorly understood classes, and with only a handful of observationally confirmed CS interaction examples, it is very difficult to understand their environments.
\end{enumerate}

Finally, we emphasize that the observational diversity among various classes of SNe exhibiting CS interaction  reflects the complex mass-loss histories of their progenitors and 
 implies that multiple mass-loss channels are at play: steady winds, binary stripping, and pre-SN eruptions. CS interaction remains the primary mechanism to constrain mass-loss histories of the progenitor in the final year to centuries.
Our findings underscore the importance of coordinated multi-wavelength observations to constrain mass-loss rates, CSM densities, and their connection to progenitor systems.

Below, we lay out an ideal roadmap to take CS interaction studies  forward in various SNe.
Towards observational efforts, early coverage (hrs--days) in ultraviolet and blue spectroscopy is required to observe shock breakout and flash ionization features, which will constrain the confined CSM size and properties. ULTRASAT \cite{BenAmi2022} will be crucial for this study. 
Radio and mm follow-up from day 1, with daily cadence for 1--2 weeks and weekly cadence spanning months to years with ALMA, VLA, GMRT, MeerKAT etc., telescopes  is crucial.  This will help break the SSA vs FFA degeneracy, provide evolution of mass-loss rate, CSM density profile, and shock speeds. The upcoming ngVLA and SKA are likely to revolutionize this field.
In X-ray bands, soft to hard X-ray follow-up from early to late time is important. \ep{} is already changing this field by detecting early soft X-ray emission in fast-evolving SNe \cite{Srinivasaragavan2025}. \swift{} can provide the required cadence in bright enough SNe to enable light curve evolution studies.  \chandra, \xmm, and \nustar{} or similar telescopes are needed for X-ray spectroscopy, to separate thermal and non-thermal components, constrain the column density, and the shock temperatures. The unprecedented sensitivity and resolution of upcoming missions like {\em NewAthena} will be instrumental.
Mid-IR spectroscopy/photometry with
 JWST/MIRI, along with ground-based NIR observations, are instrumental in
catching CDS dust formation and separating them out from different components, such as ejecta dust, dust echoes, as well as measuring their dust properties \cite{Shahbandeh25}.
Optical spectropolarimetry  will constrain the geometry of the CSM, such as equatorial rings, clumps, bipolar CSM shells, etc.
High-resolution radio imaging (VLBI) of nearby SNe will directly measure the expansion speeds, asymmetry, and equipartition parameters.
Continued monitoring and modeling — particularly of late-time radio and X-ray emission — are crucial for unveiling the final evolutionary stages of massive stars and for linking transient properties to pre-explosion environments.
Multiwaveband modeling is of crucial importance, and models should include clumping, inhomogeneities, asymmetries, dust radiative transfer, and electron scatterings.




\vspace{6pt}

\dataavailability{All the data are either published or available in archives. We encourage reader to reach out to us for any specific request which they cannot find online.}

\acknowledgments{Author thanks referees for very useful comments, which improved the manuscript tremendously. 
Author also thanks Sahana Kumar, Raphael Baer-Way, Maryam Modjaz and Roger Chevalier for useful discussions. Author acknowledges useful discussions over the years with NCRA-TIFR, Stockholm University, NRAO and University of Virginia colleagues.
This work is supported by Chandra grants GO3-24056X, DD3-24141X and GO4-25044X. The National Radio Astronomy Observatory is a facility of the National Science Foundation operated under cooperative agreement by Associated Universities, Inc. This research has made use of data obtained from the Chandra Data Archive provided by the Chandra X-ray Center (CXC).
We thank the staff of the GMRT that made these observations possible. GMRT is run by the National Centre for Radio Astrophysics of the Tata Institute of Fundamental Research. }

\conflictsofinterest{The authors declare no conflicts of interest.} 



\abbreviations{Abbreviations}{
The following abbreviations are used in this manuscript:
\\

\noindent 
\begin{tabular}{@{}ll}
ATCA & Australia Telescope Compact Array\\
BSG & Blue Super Giant\\
CCSN & Core collapse supernova\\
CD & Contact discontinuity\\
CDS & Cold-dense shell\\
CS & Circumstellar\\
CSM & Circumstellar Medium\\
ECSN & Electron-capture Supernova\\
FS & Forward shock\\
GMRT & Giant Metrewave Radio Telescope\\
GRB & Gamma-ray burst\\
IC & inverse Compton \\
IR & Infrared\\
ISM & Interstellar medium\\
LBV & Luminous blue variable\\
MERLIN & Multi-Element Radio
Linked Interferometer Network\\
MRP & Main-sequence radio-pulse emitter\\
PISN & Pair-Instability Supernova \\
PPISN & Pulsational Pair-Instability Supernova \\
PTF & Palomer Transient Factory\\
RS & Reverse shock\\
RSG & Red Super Giant\\
SESN & Stripped Envelope supernova\\
SLSN & Superluminous supernova\\
SN & Supernova\\
SNR & Supernova remnant\\
SNe II & Type II supernovae \\
SNe I & Type I supernovae \\
SNe Ia & Type Ia supernovae \\
SNe Ia-CSM & Type Ia-CSM supernovae \\
SNe IIP & Type IIP supernovae \\
SNe IIL & Type IIL supernovae \\
SNe IIb & Type IIb supernovae \\
SNe IIn & Type IIn supernovae \\
SNe Ib & Type Ib supernovae \\
SNe Ibn & Type Ibn supernovae \\
SNe Ic & Type Ic supernovae \\
SNe Icn & Type Icn supernovae \\
SNe Ic-BL & Type Ic-BL supernovae \\
UV & Ultraviolet\\
WD & While dwarf\\
WR & Wolf-Rayet\\
VLA & Karl G. Jansky Very Large Array\\
YSG & Yellow Super Giant\\
ZTF & Zwicky Transient Facility
\end{tabular}
}

\begin{adjustwidth}{-\extralength}{0cm}

\reftitle{References}


 \bibliography{ref}

%


\PublishersNote{}

\end{adjustwidth}
\end{document}